\documentclass[conference]{IEEEtran}
\ifCLASSINFOpdf
\else
\fi

\usepackage[utf8]{inputenc}
\usepackage{amssymb}
\usepackage{amsfonts}
\usepackage[dvips]{graphicx}
\usepackage{times}
\usepackage[]{cite}
\usepackage{amsmath}
\usepackage{array}
\usepackage{amssymb}

\usepackage{stfloats}
\usepackage{slashbox}
\usepackage{graphicx}
\usepackage{footnote}
\usepackage{booktabs}
\usepackage{array}
\usepackage{subeqnarray}
\usepackage{cases}
\usepackage{threeparttable}
\usepackage{color}
\usepackage{nomencl}
\makenomenclature
\usepackage[numbers,sort&compress]{natbib}
\IEEEoverridecommandlockouts
\newtheorem{theorem}{Theorem}
\newtheorem{remark}{Remark}
\newtheorem{proposition}{Proposition}
\newtheorem{lemma}{Lemma}

\newtheorem{corollary}{Corollary}
\usepackage{flafter}
\usepackage{subfigure}
\usepackage{float}
\usepackage{supertabular}
\usepackage{caption}
\usepackage{url}
\allowdisplaybreaks[4]
\usepackage[ruled,vlined,lined,ruled,linesnumbered]{algorithm2e}
\SetKwProg{Fn}{Function}{}{end}\SetKwFunction{FRecurs}{FnRecursive}%
\SetAlgoLongEnd
\SetKwComment{Comment}{$\triangleright$\ }{}

\newcommand{\bm}[1]{\mbox{\boldmath{$#1$}}}

\DeclareMathOperator*{\argmin}{arg\,min}
\usepackage{subfigure}
\usepackage{stfloats}
\usepackage{etoolbox}
\renewcommand\nomgroup[1]{%
  \item[\bfseries
  \ifstrequal{#1}{P}{Physics Constants}{%
  \ifstrequal{#1}{N}{Number Sets}{%
  \ifstrequal{#1}{O}{Other Symbols}{}}}%
]}

\begin{document}

\title{Privacy-Preserving Distributed Optimal Power Flow with Partially Homomorphic Encryption}

\IEEEoverridecommandlockouts
\author{\IEEEauthorblockN{Tong~Wu,~\IEEEmembership{Student Member,~IEEE}, Changhong~Zhao,~\IEEEmembership{Member,~IEEE} and Ying-Jun Angela Zhang, ~\IEEEmembership{Fellow,~IEEE}} \\

\thanks{T. Wu, C. Zhao, Y.-J. A. Zhang are with the Department of Information Engineering, The Chinese University of Hong Kong, Shatin, New Territories, Hong Kong (Email: \{wt017, chzhao, yjzhang\}@ie.cuhk.edu.hk).}

\thanks{This work was supported in part by General Research Funding (Project number 14208017,  Project number 14201920) and Early Career Award No. 24210220 from the Research Grants Council of Hong Kong, CUHK Research Incentive Scheme No. 4442755 and   CUHK Direct Grant No. 4055128.}
 }

\maketitle
\newcommand{\norm}[1]{\left\lVert#1\right\rVert}
\newcommand*\abs[1]{\lvert#1\rvert}
\vspace{-1.5cm}
\begin{abstract}
Distribution grid agents are obliged to exchange and disclose their states explicitly to neighboring regions to enable distributed optimal power flow dispatch. However, the states contain sensitive information of individual agents, such as voltage and current measurements. {These measurements can be inferred by adversaries, such as other participating agents or eavesdroppers, leading to the privacy leakage problem. }
  To address the issue, we propose a privacy-preserving distributed optimal power flow (OPF) algorithm based on partially homomorphic encryption (PHE).  First of all, we exploit the alternating direction method of multipliers (ADMM)  to solve the OPF in a distributed fashion. In this way, the dual update of ADMM can be encrypted by PHE. We further relax the augmented term of the primal update of ADMM with the $\ell_1$-norm regularization. In addition, we  transform the relaxed ADMM with the $\ell_1$-norm regularization to a semidefinite program (SDP), and prove that this transformation is  exact. The SDP can be solved locally with only the sign messages from neighboring agents, which preserves the privacy of the primal update. At last, we strictly prove the privacy preservation guarantee of the proposed algorithm.  {Numerical case studies validate the effectiveness and exactness of the proposed approach. In particular, the case studies show  that the encrypted messages cannot be inferred by adversaries. Besides, the proposed algorithm obtains the solutions that are very close to the global optimum, and converges much faster  compared to competing alternatives. }

\end{abstract}

\section*{Nomenclature}

\textit{Sets}
\addcontentsline{toc}{section}{Nomenclature}
\begin{IEEEdescription}[\IEEEsetlabelwidth{$V_1,V_2,V_3$}]\small
\item[$\mathcal{N}$] The set of all buses.
\item[$\mathcal{M}$] The set of all edges.
\item[$\mathcal{N}^{g}$] The set of generation buses.
\item[$\mathcal{R}_{r}$] The set of buses assigned to region $r$.
\item[$\mathcal{B}_{r}$] The joint set including the buses in $\mathcal{R}_r$ and the buses duplicated from the neighboring regions that are directly connected to the buses in $\mathcal{R}_r$.
\item[$\mathcal{N}^{\delta(r)}$]  The set of neighboring regions for region $r$.
\item[$\mathcal{F}_r^{\mathbb{C}}$]  The feasible set of the OPF problem in region $r$ in the complex domain.
\item[$\mathcal{F}_r^{\mathbb{R}}$]  The feasible set of the OPF problem in region $r$ in the real domain.
\end{IEEEdescription}

\textit{Abbreviation}
\addcontentsline{toc}{section}{Nomenclature}
\begin{IEEEdescription}[\IEEEsetlabelwidth{$V_1,V_2,V_3$}]\small
\item[ADMM] Alternating direction method of multipliers.
\item[OPF] Optimal Power Flow.
\item[DOPF] Distributed Optimal Power Flow.
\item[PHE] Partially homomorphic encryption.
\item[SDP] Semidefinite Program.
\item[PPOPF] Privacy-Preserving Distributed Optimal Power Flow.
\end{IEEEdescription}

\textit{Variables}
\addcontentsline{toc}{section}{Nomenclature}
\begin{IEEEdescription}[\IEEEsetlabelwidth{$V_1,V_2,V_3$}]
\item[$P_{G_i}$] The active  power by generators at bus $i$.
\item[$Q_{G_i}$] The reactive power by generators at bus $i$.
\item[$V_i$] The complex voltage at bus $i$.
\item[$\mathbf{v}$] The vector of  complex bus voltages.
\item[$\mathbf{W}$] The square matrix of $\mathbf{v}$.
\item[$\mathbf{X}$] The real part  of $\mathbf{W}$.
\item[$\mathbf{Z}$] The imaginary part  of $\mathbf{W}$.
\item[$\mathbf{W}^r$] The sub-matrix of $\mathbf{W}$ restricted to $\mathcal{B}_{r}$.
\item[$\mathbf{X}^r$] The  sub-matrix  of $\mathbf{X}$ restricted to $\mathcal{B}_{r}$.
\item[$\mathbf{Z}^r$] The sub-matrix   of $\mathbf{Z}$ restricted to $\mathcal{B}_{r}$.
\item[$\mathbf{W}^r_l$] The sub-matrix of $\mathbf{W}$ restricted to $\mathcal{B}_{r}\cap \mathcal{B}_{l}$ (variables for region $r$).
\item[$\mathbf{X}^r_l$] The  sub-matrix  of $\mathbf{X}$ restricted to $\mathcal{B}_{r}\cap \mathcal{B}_{l}$ (variables for region $r$).
\item[$\mathbf{Z}^r_l$] The sub-matrix   of $\mathbf{Z}$ restricted to $\mathcal{B}_{r}\cap \mathcal{B}_{l}$. (variables for region $r$)
\item[$\mathbf{\Gamma}^r_l$] The dual variable matrix corresponding to   $\mathbf{X}^r_l$.
\item[$\mathbf{\Lambda}^r_l$] The dual variable matrix corresponding to   $\mathbf{Z}^r_l$.
\item[$\mathbf{X}^r_{l,o}$] An element  of $\mathbf{X}^r_l$.
\item[$\mathbf{Z}^r_{l,o}$] An element  of $\mathbf{Z}^r_l$.
\item[$X$] A block diagonal matrix, defined in \eqref{eq42}-\eqref{bigx}. 
\end{IEEEdescription}

\textit{Constants}
\addcontentsline{toc}{section}{Nomenclature}
\begin{IEEEdescription}[\IEEEsetlabelwidth{$V_1, V_2, V_3, $}]
\item[\smash{\begin{IEEEeqnarraybox*}[][t]{l}
{a}_j,{b}_j,\\
\hphantom{V_1,{}}{c}_j
\end{IEEEeqnarraybox*}}] Coefficients of the quadratic production cost function of unit $j$.
\item[$\mathbf{Y}_i,\mathbf{\bar{Y}}_i, \mathbf{M}_i$] The constructed matrices based on $Y$ at bus $i$.
\item[$P_{D_i}$] The active  power demand at bus ${i}$.
\item[$Q_{D_i}$] The reactive  power demand at bus ${i}$.
\item[$Y$] Nodal admittance matrix.
\item[$Y_{ij}$]  The ($i,j$)th element of $Y$. 
\item[$\overline{P}_{G_i}$] The upper bound of  ${P}_{G_i}$.
\item[$\underline{P}_{G_i}$] The lower bound of  ${P}_{G_i}$.
\item[$\overline{Q}_{G_i}$] The upper bound of  ${Q}_{G_i}$.
\item[$\underline{Q}_{G_i}$] The lower bound of  ${Q}_{G_i}$.
\item[${\overline{V}_i}$] The upper bound of voltage magnitude $|V_{i}|$.
\item[${\underline{V}_i}$] The lower bound of voltage magnitude $|V_{i}|$
\item[$\mathbf{W}^l_r$] The communication message that is a sub-matrix of $\mathbf{W}$ restricted to $\mathcal{B}_{r}\cap \mathcal{B}_{l}$ (constants  for agent $r$).
\item[$\mathbf{X}^l_r$]  The communication message that is a   sub-matrix  of $\mathbf{X}$ restricted to $\mathcal{B}_{r}\cap \mathcal{B}_{l}$ (constants for agent $r$).
\item[$\mathbf{Z}^l_r$]  The communication message that is a  sub-matrix   of $\mathbf{Z}$ restricted to $\mathcal{B}_{r}\cap \mathcal{B}_{l}$ (constants for agent $r$).
\item[$\mathbf{G}_i^r$, $\mathbf{\bar{G}}_i^r$] The real part of $\mathbf{Y}_i,\mathbf{\bar{Y}}_i$.
\item[$\mathbf{B}_i^r$, $\mathbf{\bar{B}}_i^r$] The imaginary part of $\mathbf{Y}_i,\mathbf{\bar{Y}}_i$.
\end{IEEEdescription}

\section{Introduction}
\subsection{Background and Motivation}
With the increasing observability of power systems, the system operators carry out advanced operational practices to steer the system  towards  an optimal power flow (OPF) solution \cite{lavaei2011zero, wu2019vsc}.  In particular,  when the synchronized measurements from different buses are transferred to the system operator, the sensitive information, such as voltage and current measurements, is likely to expose distribution grid buses to privacy breaches \cite{dvorkin2019differentially}. Several studies have shown that the high-resolution measurements of OPF variables can be used by adversaries to infer the users' energy consumption patterns and  types of appliances \cite{wang2020tsg,liu2012cyber}.

To reduce the global communication of different buses, distributed algorithms have been proposed to solve OPF, where only boundary variables are shared among agents that act as independent system operators in  different regions \cite{peng2016distributed, dall2013distributed, erseghe2014distributed, zhang2014optimal, lin2019decentralized, zhao2019distributed}. In these works, one classical approach is  alternating direction method of multipliers (ADMM) \cite{peng2016distributed, dall2013distributed, erseghe2014distributed}. In particular, Reference \cite{dall2013distributed}  applied the ADMM to  the semidefinite program (SDP) relaxation of OPF problem. Then, \cite{erseghe2014distributed} solved the general form of OPF problems using ADMM in a distributed way, where the complex voltages on the boundary buses are shared with neighboring agents. Similarly, the OPF problem has been relaxed as a second order cone program (SOCP), and then solved in a distributed way by  ADMM \cite{peng2016distributed}. Noticeably, the complex power flow measurements in \cite{peng2016distributed} are transferred to neighboring agents. Another distributed approach is  dual decomposition \cite{lam2012distributed, zhao2019distributed}. Reference \cite{lam2012distributed} proposed  a dual algorithm to coordinate the subproblems decomposed from the SDP relaxation of OPF problem. In \cite{zhao2019distributed}, the decomposed subproblem  can be solved  with a closed-form solution. {Meanwhile, the solutions of the subproblems need to be communicated between each pair of agents iteratively, resulting in large  privacy leakage.}  Other methods include the  multiplier methods \cite{zhang2014optimal, lin2019decentralized}, where the buses are required to exchange their states directly in the communication network. 


To achieve or approximate the optimal solution in the above distributed algorithms,  agents inevitably need to share their individual information with their neighbors, which leads to serious privacy leakage problems. {For example, the direct communication among buses gives adversaries or eavesdroppers a chance to launch the cyber attack on the grid system effectively \cite{musleh2019survey}.} In particular, adversaries are able to design the optimal attack to increase the generation costs or disturb the electricity market \cite{liu2012cyber}. Therefore,  it is  desirable to design privacy-preserving distributed OPF (DOPF) algorithms.

\subsection{Related Works}
Recently, some privacy-preserving methods have been proposed to solve distributed optimization problems. These  methods are broadly classified into two categories, namely differential-privacy methods \cite{han2016differentially, dvorkin2020differentially, mak2019privacy} and homomorphic encryption methods \cite{lu2018privacy, zhang2018enabling, zhang2018admm}. In particular, differential-privacy methods  introduce a random perturbation to the shared messages to protect an agent's privacy \cite{han2016differentially}. In power systems, \cite{dvorkin2020differentially}  optimized OPF variables as affine functions of the random noise. {In addition,  \cite{mak2019privacy}  introduced the OPF Load Indistinguishability (OLI) problem, which guarantees load data privacy while achieving a feasible and near optimal energy dispatch. In \cite{dvorkin2020differentially}, a differential privacy method has been proposed to protect OPF variables. Note that there is an inevitable trade-off between optimality and privacy in differential-privacy methods, due to the random perturbation added to the shared messages. }

As to the homomorphic encryption methods, \cite{lu2018privacy} studied how a system operator and a set of agents securely  execute a distributed projected gradient-based algorithm. In \cite{zhang2018enabling, zhang2018admm}, the projected gradient-based algorithm and ADMM algorithm were incorporated with the partially homomorphic encryption scheme  (PHE) to facilitate the privacy-preserving distributed optimization. {In \cite{zhang2020secure}, a PHE-based method has been proposed to estimate states securely. In addition, \cite{alexandru2020cloud} applies the PHE to the cloud-based quadratic optimization problem.
However, these methods are applicable only when the optimization problems are unconstrained or a closed-form primal update is available. Therefore, the above methods are not applicable to the OPF problems directly.}


\subsection{Contributions}

To address the challenges mentioned above, we are motivated to develop a privacy-preserving DOPF method based on partially homomorphic encryption.  To enable the cryptographic techniques in DOPF, we first apply a sequence of strong SDP relaxations to the OPF problem. Then, we decouple the centralized  SDP into multiple small-scale SDPs with boundary constraints. In this way,  the ADMM method is employed to solve multiple SDPs in a distributed fashion. In particular, we protect the privacy of both primal and dual updates of ADMM based on PHE. 
%
%
Our main contributions are summarized as follows.
\begin{itemize}
	\item We develop a privacy-preserving distributed optimal power flow method based on PHE. To the best of our knowledge, this is the first time that cryptographic techniques are incorporated in the distributed AC optimal power flow problem. {Compared with differential-privacy based optimization methods,  our approach obtains the solutions that are very close to the global optimum, and converges faster. In addition, our approach can preserve the privacy of   gradients and intermediate primal states.}
	\item  To solve the primal update problem securely, we relax the augmented terms in ADMM by   the $\ell_1$-norm regularization. We further transform  the new primal update problem as an SDP problem, which only shares the sign information with other agents. In addition, the privacy of the dual update of ADMM is protected by PHE with random penalty parameters. {In comparison to the traditional ADMM that exposes agents' intermediate states to privacy breaches, the proposed ADMM protects the privacy of both primal and dual updates of ADMM.} The difference between the traditional ADMM  and proposed method is summarized in Table \ref{table1}.

		\item  We rigorously prove the privacy preservation of the proposed algorithm with two commonly encountered types of adversaries.  For honest-but-curious agents, our analysis shows that the privacy information of neighboring agents cannot be inferred from the shared messages. Likewise, we  prove that external eavesdroppers cannot infer the privacy of exchanged information.   {In contrast, other PHE-based algorithms cannot protect the primal   update of ADMM if the primal  problem is constrained.}
\end{itemize}
    \begin{table}[!h]\footnotesize
\center
\begin{threeparttable}
\caption{Traditional ADMM  VS  Privacy-Preserving ADMM}
\label{table1}
\centering
\begin{tabular}{c |c}
\toprule
Traditional ADMM& Privacy-Preserving ADMM\\
\midrule
Frobenius norm & $\ell_1$-norm \\
Synchronized  & Asynchronized \\
Privacy leakage& Privacy preservation\\
\bottomrule
\end{tabular}
\end{threeparttable}
\vspace{-0.5cm}
\end{table}

\subsection{Organization and Overview}
The rest of the paper is organized as follows. In Section II, we introduce the preliminaries of the OPF problem and Paillier cryptosystem. In Section III, we present the privacy-preserving distributed optimal power flow method. In Section IV, we strictly prove the privacy preservation guarantee of the proposed algorithm. The case studies  are given in Section V. Finally, we conclude the paper in Section VI.

\begin{figure}[!htb]
\centering
\includegraphics[width=3.4in]{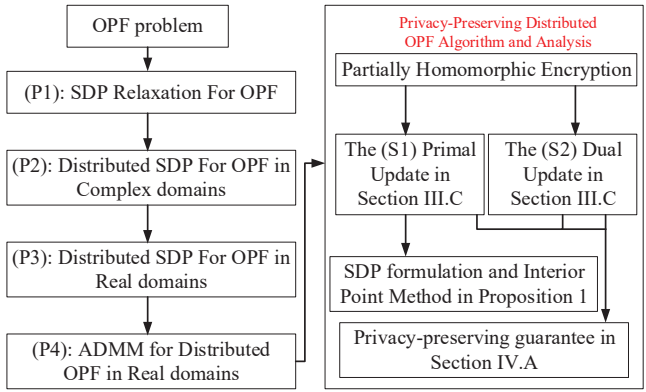}
\caption{The overview of  the proposed method.}\label{overview}
\vspace{-0.4cm}
\end{figure}  

{The schematic overview of the methodology in the paper is shown in Fig. \ref{overview}. We first review the OPF problem (P1), and then transform it into the SDP problem both in complex domains and real domains, i.e., Problem (P2) and Problem (P3). Furthermore, we equivalently transform (P3) into the distributed formulation, and further employ the ADMM method for the DOPF problem in (P4). The ADMM method is relaxed by the  $\ell_1$-norm regularization. As such, we develop the (S1) primal update and the (S2) dual update in the relaxed ADMM method. In the process, both the (S1) primal update and the (S2) dual update are encrypted by the Partially Homomorphic Encryption scheme. In particular, the (S1) primal update is formulated as an SDP problem, and then solved by the interior-point method, elaborated in Proposition 1 and Appendix. At last, we provide the mathematical proof for the privacy-preserving guarantee in Section IV.A.}

\section{Preliminaries}
\subsection{Definition}
It is worth noticing that privacy has different meanings for different applications. For example, privacy has been defined as the non-disclosure of agent's states \cite{lu2018privacy, zhang2018enabling}, gradients or sub-gradients \cite{dvorkin2020differentially, mak2019privacy}.  In this paper, we define privacy as the non-disclosure  of agents' intermediate states and gradients of the objective functions. {Noticeably, agents' intermediate states are feasible solutions to power flow equations in corresponding regions. Besides, power flow equations can be regarded as a mapping function from intermediate states to the system information.
Therefore, the parameters and topologies of subregions can be inferred by data mining tools from a long-term observation of agents' intermediate states \cite{yu2017patopa}.  In general, distributed algorithms usually  take multiple iterations to converge, which generates massive intermediate states that enable adversaries to infer the grid models effectively. Likewise, the objective parameters can be inferred from  agents' intermediate states and gradients of the objective functions in the same way.
 In addition,  undetectable attacks can be designed by  partial observations of  agents' intermediate states \cite{liu2016false}. During this process, to make an attack undetectable, adversaries should successively inject false data into agents' intermediate states, since only injecting false data into the final states can be easily detected. 
 Therefore, if unprotected, agents' intermediate states could be obtained by adversaries to attack the power grids or disturb the electricity market.}

We also define two kinds of adversaries: honest-but-curious agents and  external eavesdroppers \cite{zhang2018admm}. Honest-but-curious agents  are the agents that follow all protocol steps but are curious and collect all the intermediate states  of other participating agents. External eavesdroppers are adversaries who steal information through  eavesdropping all the communication channels and exchanged messages between agents. {The differences between honest-but-curious agents and  external eavesdroppers are: 
\begin{itemize}
	\item  honest-but-curious agents only obtain information from the neighboring agents, while external eavesdroppers obtain all agents' shared information;
	\item honest-but-curious agents can decrypt the shared encrypted information, but  external eavesdroppers cannot.
\end{itemize}}
Preserving the privacy of agents' intermediate states can prevent eavesdroppers from inferring any information in optimization.

\subsection{Optimal Power Flow}
We represent the power network by a graph $\mathcal{G}(\mathcal{N}, \mathcal{M})$, with  vertex set $\mathcal{N}$ and edge set $\mathcal{M}$.  We consider the following OPF optimization problem \cite{wu2021distributed}:
\begin{subequations}\label{eq:op1}
\begin{alignat}{2}
&\min_{P_{G_i}, Q_{G_i}, V_i}&&	C(P_{G_i})=\sum_{i=1}^{\abs{\mathcal{N}^g}}a_iP_{G_i}^2+b_iP_{G_i}+c_i,\label{eq:op1_obj}\\
&\text{s.t.}&& P_{G_i}-P_{D_i}= \text{Re}\{V_i\sum_{j\in \delta(i)}( Y_{ij}V_j)^*\},i\in \mathcal{N},\label{eq:op1_cst1}\\
&&& Q_{G_i}-Q_{D_i}= \text{Im}\{V_i\sum_{j\in \delta(i)}( Y_{ij}V_j)^*\},i\in \mathcal{N},\label{eq:op1_cst2}\\
&&& \underline{V_{i}}\le \abs{V_{i}} \le \overline{V_{i}},i\in \mathcal{N},\label{eq:op1_cst3}\\
&&& \underline{P}_{G_i}\le P_{G_i} \le \overline{P}_{G_i},i\in \mathcal{N},\label{eq:op1_cst4}\\
&&& \underline{Q}_{G_i}\le Q_{G_i} \le \overline{Q}_{G_i},i\in \mathcal{N}.\label{eq:op1_cst5}
\end{alignat}
\end{subequations}
where $Y$ be the nodal admittance matrix and $\delta(i)$ is the set of neighboring buses of node $i$. 
The objective function \eqref{eq:op1_obj} describes the fuel cost. $a_i$, $b_i$ and $c_i$ are nonnegative  coefficients. Note that $\mathcal{N}^g$ is the set of generator buses and $\abs{\mathcal{N}^g}$ is its cardinality. $V_{i}$ is the complex voltage on bus $i$ with  magnitude $\abs{V_{i}}$ and  phase $\abs{\theta_{i}}$. 
The constraints in \eqref{eq:op1_cst1} and  \eqref{eq:op1_cst2} describe the power flow equations on each bus $i$. The   active power output $P_{G_i}$ and  reactive power output $Q_{G_i}$ of generator $i$ and the voltage magnitude $\abs{V_i}$  are bounded in Eqs. \eqref{eq:op1_cst3}-\eqref{eq:op1_cst5} by constant bounds $\underline{P}_{G_i}$, $\overline{P}_{G_i}$, $\underline{Q}_{G_i}$, $\overline{Q}_{G_i}$, $\underline{V_{i}}$ and $\overline{V_{i}}$.  In addition,  we enforce  $\underline{P}_{G_i}=0$, $\overline{P}_{G_i}=0$, $\underline{Q}_{G_i}=0$ and $\overline{Q}_{G_i}=0$ for non-generator buses $\mathcal{N}/\mathcal{N}^g$.  Problem \eqref{eq:op1} is a non-convex  problem because of the nonconvexity of constraints \eqref{eq:op1_cst1}-\eqref{eq:op1_cst2}.

\subsection{Partially Homomorphic Encryption}
In this subsection, we introduce cryptosystem, which  will be used to enable privacy-preserving DOPF in the following section. The Paillier cryptosystem is a public-key system consisting of  three parts, i.e., key generation, encryption and decryption \cite{paillier1999public}. In particular, we have the public key and private key in the key generation stage. The public key are disseminated to all agents  to encrypt  messages. The private key is only known to one agent or the {system operator} and used to decrypt messages. In particular, both the private key and the public key  are  time-varying. {As shown in Algorithm \ref{alg:x},  the three parts of Paillier cryptosystem are implemented by three functions, i.e., \textbf{Keygen}(), $c=\mathcal{E}(m)$ and $m=\mathcal{D}(c)$.

Noticeably, the Paillier system is  additively homomorphic. In particular, the ciphertext of $m_1+m_2$ can be obtained from the ciphertexts of $m_1$ and $m_2$ directly. We have
\begin{align}
	\mathcal{E}(m_1)*\mathcal{E}(m_2)=\mathcal{E}(m_1+m_2),\\
	\mathcal{E}(m)^k=\mathcal{E}(km), k\in \mathbb{Z}^{+},
\end{align}
where $m_1$ and $m_2$ are the original messages, and $c_1$ and $c_2$ are the ciphertext of $m_1$ and $m_2$, respectively.}

   \begin{algorithm} [!htb]\small
       \caption{Paillier cryptosystem}
    \label{alg:x} 
  \Fn(){\textbf{Keygen} ()}{
      \KwOut{Public key $k_p$: ($n, g$), Private key $k_s$: ($\lambda, \mu$)} 
        Choose two large prime numbers $p$ and $q$ of equal  bit-length and compute $n=p\cdot q$\;
        $g\longleftarrow n+1$\;
        $\lambda=\phi(n)=(p-1)\cdot (q-1)$, where $\phi(\cdot)$ denotes the Euler's  totient function\;
        $\mu=\phi^{-1}(n)$ mod $n$, which is the modular multiplicative inverse of $\phi(n)$\;
  }	
  \Fn(){$\mathcal{E}$(m)}{
      \KwOut{Ciphertext $c$} 
      Choose a random $r\in \mathbb{Z}_n^*=\{z|z\in \mathbb{Z}, 0\le z < n, $ $gcd(z,n)=1\}$\;
      Generate the ciphertext  by $c=g^m\cdot r^n$ mod $n^2$, where $m\in \mathbb{Z}_n=\{z|z\in \mathbb{Z}, 0\le z < n\}$, $c\in \mathbb{Z}_{n^2}^*$\; 
  }	  
    \Fn(){$\mathcal{D}$(c)}{
      \KwOut{Message $m$} 
      Define the integer division function $L(\mu)=\frac{\mu-1}{n}$\;
      Generate the plaintext as $m=L(c^{\lambda} \text{ mod } n^2)\cdot \mu \text{ mod } n$\; 
  }	
   \end{algorithm}

\section{Privacy-Preserving Distributed Optimal Power Flow}
\subsection{Convex Relaxation}
To deal with the nonconvexity of Problem \eqref{eq:op1}, we apply SDP relaxation in this subsection. 

Let $\mathbf{v}$ define the vector of  complex bus voltages $\mathbf{v}=(\abs{V_1}\angle \theta_1, \cdots, \abs{V_N}\angle \theta_N)$, where $N$ denotes the number of buses.
We define variable matrix $\mathbf{W}=\mathbf{v}\mathbf{v}^H$. We use the notation $\text{Tr}\{\mathbf{X}\}$ to represent the trace of an arbitrary square matrix $\mathbf{X}$.
%
%
Recall that $Y$ is the admittance matrix. For $i\in \mathcal{N}$, $e_i$ is the $i$th basis vector in $\mathbb{R}^{N}$, $e_i^T$ is its transpose, and $Y_i=e_ie^T_iY$. We define $\mathbf{Y}_i= {1}/{2}(Y_i^H+Y_i)$, $\bar{\mathbf{Y}}_i= {1}/{2} \sqrt{-1}(Y_i^H-Y_i)$ and ${\mathbf{M}}_i=e_ie^T_i$. Then, we have $P_{G_i}=\text{Tr}\{\mathbf{Y}_i\mathbf{W}\}+P_{D_i}$.  

The active and reactive power balance equations  Eq. \eqref{eq:op1_cst1} and Eq. \eqref{eq:op1_cst2}  can be combined with constraints \eqref{eq:op1_cst4} and \eqref{eq:op1_cst5} as
\begin{equation}
\begin{split}
	\underline{P}_{G_i}\le P_{G_i}=\text{Tr}\{\mathbf{Y}_i\mathbf{W}\}+P_{D_i} \le \overline{P}_{G_i},\label{eq:sdp2_cst1}
\end{split}
\end{equation}
\begin{equation}
\begin{split}
	\underline{Q}_{G_i}\le Q_{G_i}= \text{Tr}\{\bar{\mathbf{Y}}_i \mathbf{W}\}+Q_{D_i} \le \overline{Q}_{G_i},\label{eq:sdp2_cst2}
\end{split}
\end{equation}

Moreover, Eq. \eqref{eq:op1_cst3} can be transformed to
\begin{equation}
\begin{split}
	(\underline{V_{i}} )^2\le \text{Tr}\{{\mathbf{M}}_i\mathbf{W}\} \le (\overline{V_{i}} )^2\label{eq:sdp2_cst3}.
\end{split}
\end{equation}

As such, we can write an equivalent form of Problem \eqref{eq:op1} as follows
\begin{subequations}\label{eq:op2}
\begin{alignat}{2}
(\text{P1}):\phantom{xx}&\min&\phantom{xxx}&	 C(P_{G_i})=\sum_{i=1}^{\abs{\mathcal{N}^g}}a_iP_{G_i}^2+b_iP_{G_i}+c_i,\\
&\text{s.t.}&\phantom{xx}& \text{\eqref{eq:sdp2_cst1}-\eqref{eq:sdp2_cst3}}, \label{eq:opt_f1}\\
&\text{}&\phantom{xx}&\mathbf{W} \text{ is hermitian, }  \mathbf{W} \succeq 0, \label{eq:psd}\\
&\text{}&\phantom{xx}& \text{rank}(\mathbf{W})=1,\label{eq:opt_f2}
\end{alignat}
\end{subequations}
where the rank-1 constraint in \eqref{eq:opt_f2} makes the problem  nonconvex. A convex SDP relaxation of \eqref{eq:op2} is obtained by removing the rank constraint \eqref{eq:opt_f2}. By Theorem 9 of \cite{low2014convex}, when the cost function is convex and the network is a tree, the  SDP relaxation is exact under mild technical conditions. 

\subsection{Distributed Formulation}
%
Let $R$ be the total number of regions and $\mathcal{R}_r$ be the set of buses assigned to region $r$ with $\mathcal{R}_r \cap \mathcal{R}_l=\emptyset$, $\forall l \neq r$ and $\sum_{r=1}^R\abs{\mathcal{R}_r}=\abs{\mathcal{N}}$. Let $\mathcal{B}_r$ denote the joint set including the buses in $\mathcal{R}_r$ and the buses duplicated from the neighboring regions that are directly connected to the buses in $\mathcal{R}_r$. Here,  the set of neighboring regions for region $r$ can be expressed   as $\mathcal{N}^{\delta(r)}=\{l|\mathcal{B}_r \cup \mathcal{B}_l \neq 0 \}$. 
Finally, we stack the complex voltages of the nodes in $\mathcal{B}_r$ as ${\mathbf{v}}^r$, i.e., $\{V_i\}_{i\in \mathcal{B}_r}$. 

In addition,  $\mathbf{W}=\mathbf{X}+\sqrt{-1}\mathbf{Z}$, where $\mathbf{X}$ is the real part of $\mathbf{W}$ and $\mathbf{Z}$ is the imaginary part of $\mathbf{W}$.
Moreover, we define $\mathbf{Y}_i^r$, $\bar{\mathbf{Y}}_i^r$, $\mathbf{M}_i^r$, $\mathbf{W}^r$, $\mathbf{X}^r$ and $\mathbf{Z}^r$ as the sub-matrices of $\mathbf{Y}_i$, $\bar{\mathbf{Y}}_i$, $\mathbf{M}_i$, $\mathbf{W}$, $\mathbf{X}$ and $\mathbf{Z}$, respectively. In particular, these sub-matrices  are formed by extracting rows and columns corresponding to the nodes in $\mathcal{B}_r$. 
%
%
Next,  $\mathcal{P}_{rl}$  indexes voltages at the buses shared by  $\mathcal{B}_r$ and $\mathcal{B}_l$. For example, if region $1$ and region $2$ share nodes $n=3$ and $n=4$, then $\mathcal{P}_{12}$ indexes the voltage $\{V_3\}$ and $\{V_4\}$. In addition, let $\mathbf{W}^r_l$ denote the submatrix of $\mathbf{W}^r$, which collects the rows and columns of $\mathbf{W}^r$ corresponding to the voltages in $\mathcal{P}_{rl}$. As such, we rewrite problem (P1) (after rank relaxation)  in the following equivalent form:
\begin{subequations}
	\begin{alignat}{2}
(\text{P2}):\phantom{x}&\min&\phantom{xx}&\sum_{r=1}^RC^r(P_G^r)=\sum_{r=1}^R(\sum_{i=1}^{\abs{\mathcal{R}_r^g}} a_iP_{G_i}^2+b_iP_{G_i}+c_i),\\
&\text{s.t.}&\phantom{x}& \{\mathbf{W}^r, P_G^r, Q_G^r\}  \in \mathcal{F}_r^{\mathbb{C}}, \forall r = 1, \cdots, R, \label{eq:p21}\\
&\text{}&\phantom{x}& \mathbf{W}^r_l=\mathbf{W}^l_r, l\in \mathcal{N}^{\delta(r)}, r = 1, \cdots, R, \label{eq:bdr01}\\
&\text{}&\phantom{x}& \mathbf{W}^r\succeq 0, \forall r = 1, \cdots, R,
	\end{alignat}
\end{subequations}
where $P_G^r$ and $Q_G^r$ denote the entire vectors of $P_{G_i}$ and $Q_{G_i}$ $\forall i\in \mathcal{B}_r$ respectively, $\mathcal{F}_r^{\mathbb{C}}$ in constraint \eqref{eq:p21}  denotes the feasible set \eqref{eq:sdp2_cst1}-\eqref{eq:sdp2_cst3}   restricted to buses   $\mathcal{B}_r$. Besides,  constraint \eqref{eq:bdr01} enforces neighboring areas to consent on the entries of $\mathbf{W}^r$ and $\mathbf{W}^l$ that they have in common.  $\mathcal{R}_r^g$ denotes the set of generator buses in $\mathcal{R}_r$, and $\abs{\mathcal{R}_r^g}$ denotes the number of generator buses in $\mathcal{R}_r$. We denote the real  and imaginary part of $\mathbf{W}^r_l$ by $\mathbf{X}^r_l$ and $\mathbf{Z}^r_l$, respectively. Note that $\mathbf{W}^l_r$ is the message that is transferred from region $l$ to region $r$. 
To enforce the equivalence of $\mathbf{W}^r_l$ and $\mathbf{W}^l_r$, we have $\mathbf{X}^r_l=\mathbf{X}^l_r$ and $\mathbf{Z}^r_l=\mathbf{Z}^l_r$.  We  also denote the number of buses in $\mathcal{B}_r$ by $\bar{N}_r=\abs{\mathcal{B}_r}$ and the number of buses in $\mathcal{B}_r\cap \mathcal{B}_l$ by $\bar{N}_{rl}=\abs{\mathcal{B}_r\cap \mathcal{B}_l}$.

The main challenge to this decomposition is the positive semi-definite (PSD) constraints \eqref{eq:psd}. Indeed,   all sub-matrices $\mathbf{W}^r\succeq 0, \forall r$ does not necessarily imply $\mathbf{W}\succeq 0$. 
Luckily, as established in \cite{grone1984positive}, $\mathbf{W}^r\succeq 0$ for all region $r$ is equivalent to $\mathbf{W}\succeq 0$  if the following assumption holds.
\begin{itemize}
	\item  Every maximal clique is contained in the subgraph formed by $\mathcal{B}_r$ for at least one $r$.
\end{itemize}
 When the network is a tree, every pair of adjacent nodes connected by an edge forms a maximal clique, and thus this assumption is trivially true.

Let $\mathbf{G}_i^r$ denote the real part and $\mathbf{B}_i^r$ denote the imaginary part of $\mathbf{Y}_i^r$. Let $\bar{\mathbf{G}}_i^r$ denote the real part and $\bar{\mathbf{B}}_i^r$ denote the imaginary part of $\bar{\mathbf{Y}}_i^r$. We have
\begin{equation}
\begin{split}
\text{Tr}\{\mathbf{Y}^r_i\mathbf{W}^r\}&=\text{Tr}\{(\mathbf{G}^r_i\mathbf{X}^r-\mathbf{B}^r_i\mathbf{Z}^r)+ \sqrt{-1} (\mathbf{B}^r_i\mathbf{X}^r+\mathbf{G}^r_i\mathbf{Z}^r)\}\\
&=\text{Tr}\{(\mathbf{G}^r_i\mathbf{X}^r-\mathbf{B}^r_i\mathbf{Z}^r)\},
\end{split}
\end{equation}
where $\text{Tr}\{\sqrt{-1}(\mathbf{B}^r_i\mathbf{X}^r+\mathbf{G}^r_i\mathbf{Z}^r)\}=0$ because $P_{G_i}$ is a real number. 
%
In addition, $\mathcal{F}_r^{\mathbb{R}}$ in Eqs. \eqref{eq:sdp2_cst1}-\eqref{eq:sdp2_cst3} can be equivalently expressed as: 
\begin{subequations}\label{feasible2}
	\begin{alignat}{2}
	&\underline{P}_{G_i}\le {P}_{G_i}=  \text{Tr}\{(\mathbf{G}^r_i\mathbf{X}^r-\mathbf{B}^r_i\mathbf{Z}^r)\} +P_{D_i} \le \overline{P}_{G_i},\\
	&\underline{Q}_{G_i}\le {Q}_{G_i}= \text{Tr}\{(\bar{\mathbf{G}}^r_i\mathbf{X}^r-\bar{\mathbf{B}}^r_i\mathbf{Z}^r)\}+Q_{D_i} \le \overline{Q}_{G_i},\\
	&(\underline{V_{i}} )^2\le \text{Tr}\{{\mathbf{M}}_i^r\mathbf{X}^r\} \le (\overline{V_{i}} )^2.
	\end{alignat}
\end{subequations}

We define the matrix $\mathbf{E}_{r\mapsto l}\in \mathbb{R}^{\bar{N}_{rl}\times \bar{N}_{r}}$ as
\begin{equation}
\begin{split}
	\mathbf{E}_{r\mapsto l}=\left[\bigoplus_{i\in \mathcal{P}_{rl}} {e_i^r}^T; \right],
\end{split}
\end{equation}
where $e_i^r \in \mathbb{R}^{\bar{N}_{r}}$ and $\bigoplus$ denotes the operator of concatenating  vectors.
\begin{lemma}\label{lemma1}
We have the following relationship:
\begin{equation}
\begin{split}
	\mathbf{W}\succeq 0 \Longleftrightarrow \begin{bmatrix} \mathbf{X} & -\mathbf{Z} \\ \mathbf{Z} & \mathbf{X} \end{bmatrix}\succeq 0
\end{split}
\end{equation}
\end{lemma}
\emph{Proof:} We show the detailed proof in the Appendix.

Based on Lemma \ref{lemma1}, we write (P2) in terms of $\mathbf{X}^r, \mathbf{Z}^r$ as follows:
\begin{subequations}
	\begin{alignat}{2}
(\text{P3}):&\min&\phantom{xx}&\sum_{r=1}^RC^r(P_{G}^r)\\
&\text{s.t.}&& \{\mathbf{X}^r, \mathbf{Z}^r,  P_G^r, Q_G^r\} \in \mathcal{F}_r^{\mathbb{R}}, \forall r = 1, \cdots, R,\\
&\text{}&& \mathbf{E}_{r\mapsto l}\mathbf{X}^r\mathbf{E}_{r\mapsto l}^T=\mathbf{X}^l_r, l\in \mathcal{N}^{\delta(r)}, r = 1, \cdots, R, \label{eq:bdr02}\\
&\text{}&& \mathbf{E}_{r\mapsto l}\mathbf{Z}^r\mathbf{E}_{r\mapsto l}^T=\mathbf{Z}^l_r, l\in \mathcal{N}^{\delta(r)}, r = 1, \cdots, R, \label{eq:bdr03}\\
&\text{}&& \begin{bmatrix} \mathbf{X}^r & -\mathbf{Z}^r \\ \mathbf{Z}^r & \mathbf{X}^r \end{bmatrix}\succeq 0, \forall r = 1, \cdots, R,
	\end{alignat}
\end{subequations}
where $\mathbf{X}^r_l=\mathbf{E}_{r\mapsto l}\mathbf{X}^r\mathbf{E}_{r\mapsto l}^T$,  $\mathbf{Z}^r_l=\mathbf{E}_{r\mapsto l}\mathbf{Z}^r\mathbf{E}_{r\mapsto l}^T$, and $\mathbf{E}_{r\mapsto l}\mathbf{E}_{r\mapsto l}^T=\mathbf{I}_{r\mapsto l} \in  \mathbb{R}^{\bar{N}_{rl}\times \bar{N}_{rl}}$.

\subsection{Distributed Optimal Power Flow}

Solving (P3) directly is likely to incur infeasible solutions. In particular, when we solve (P3) for region $r$ and pass the messages to  region $l$, the additional equality constraints, i.e.,  \eqref{eq:bdr02} - \eqref{eq:bdr03}, may cause the boundary variables  described by the messages infeasible  for \eqref{eq:bdr03} in region $l$. This is because the  number of equality  constraints in each subproblem may be larger than the number of its variables or these messages do not lie in the intersection of the feasible regions of  all the subproblems.

Therefore, in this subsection, we propose a distributed algorithm to solve (P3). Instead of solving (P3) directly, we relax (P3) by utilizing the augmented partial Lagrangian. In particular, \eqref{eq:bdr02} - \eqref{eq:bdr03} in region $r$ are replaced by the augmented  Lagrangian terms in its objective.  Once the distributed algorithm converges,  optimal solutions to  the augmented partial Lagrangians of different regions   will satisfy \eqref{eq:bdr02} - \eqref{eq:bdr03} simultaneously.
We consider the partial augmented Lagrangian:
\begin{subequations}\label{Lagrangian}
	\begin{alignat}{2}
(\text{P4}):	&\min_{( \{\mathbf{X}^r\}, \{\mathbf{Z}^r\}, \{\mathbf{\Gamma}^r_l\}, \{\mathbf{\Lambda}^r_l\}, r=1, \cdots, R)}\mathcal{L}_{admm}=\sum_{r=1}^R \mathcal{L}_{admm}^r=\notag\\
	& \sum_{r=1}^R\bigg\{C^r(P_{G}^r)+\sum_{l\in \mathcal{N}^{\delta{(r)}}}\left[\text{Tr}\left((\mathbf{\Gamma}^r_l)^T (\mathbf{E}_{r\mapsto l}\mathbf{X}^r\mathbf{E}_{r\mapsto l}^T-\mathbf{X}^l_r) \right)\right.\notag\\
	&	+ \text{Tr}\left((\mathbf{\Lambda}^r_l)^T (\mathbf{E}_{r\mapsto l}\mathbf{Z}^r\mathbf{E}_{r\mapsto l}^T-\mathbf{Z}^l_r) \right)+ \notag\\
	& \frac{1}{2}\norm{\sqrt{\rho_{rl}}\circ\left(\mathbf{E}_{r\mapsto l}\mathbf{X}^r\mathbf{E}_{r\mapsto l}^T-\mathbf{X}^l_r\right) }^2_F+\notag\\
	& \left.\frac{1}{2}\norm{\sqrt{\kappa_{rl}}\circ\left(\mathbf{E}_{r\mapsto l}\mathbf{Z}^r\mathbf{E}_{r\mapsto l}^T-\mathbf{Z}^l_r\right) }^2_F\right]\bigg\}\\
	&\text{s.t.} \{\mathbf{X}^r, \mathbf{Z}^r,  P_G^r, Q_G^r\} \in \mathcal{F}_r^{\mathbb{R}}, \forall r = 1, \cdots, R,\\
	& \begin{bmatrix} \mathbf{X}^r & -\mathbf{Z}^r \\ \mathbf{Z}^r & \mathbf{X}^r \end{bmatrix}\succeq 0, \forall r = 1, \cdots, R,
	\end{alignat}
\end{subequations}
where $\rho_{rl},\kappa_{rl} \in \mathbb{R}^{\bar{N}_{rl}\times \bar{N}_{rl}}$ are matrices with positive entries, and $\circ$ denotes Hadamard product. In addition, $\{\mathbf{\Gamma}^r_l\}$ and $\{\mathbf{\Lambda}^r_l\}$ denote the multipliers associated with \eqref{eq:bdr02} and \eqref{eq:bdr03}. Note that, the gradients of $\frac{1}{2}\norm{\sqrt{\rho_{rl}}\circ\left(\mathbf{E}_{r\mapsto l}\mathbf{X}^r\mathbf{E}_{r\mapsto l}^T-\mathbf{X}^l_r\right) }^2_F$ and $\frac{1}{2}\norm{\sqrt{\kappa_{rl}}\circ\left(\mathbf{E}_{r\mapsto l}\mathbf{Z}^r\mathbf{E}_{r\mapsto l}^T-\mathbf{Z}^l_r\right) }^2_F$ require successive communication between agents in each iteration, which can disclose the states of neighboring agents. {Taking $\mathbf{X}^r$ for example,  the gradient of $\frac{1}{2}\norm{\sqrt{\rho_{rl}}\circ\left(\mathbf{E}_{r\mapsto l}\mathbf{X}^r\mathbf{E}_{r\mapsto l}^T-\mathbf{X}^l_r\right) }^2_F$ is ${{\rho_{rl}}\circ\left(\mathbf{E}_{r\mapsto l}\mathbf{X}^r\mathbf{E}_{r\mapsto l}^T-\mathbf{X}^l_r\right) }$. Observing from the gradient, it is obvious that updating $\mathbf{X}^r$ by agent $r$ in each iteration requires the values of $\mathbf{X}^l_r$ from neighboring agent $l$. As a result, the values of $\mathbf{X}^l_r$ are exposed to both honest-but-curious agents and external eavesdroppers.

To solve this problem, we approximate the augmented terms by the $\ell_1$-norm regularization.  The mechanism is motivated by the fact that the subgradients of the $\ell_1$-norm regularization terms are the sign messages. Therefore, we can introduce a third party to provide the sign messages for agent $r$ to update $\mathbf{X}^r$. In this way,  the two kinds of adversaries cannot infer any information  through  eavesdropping all the communication channels and exchanged messages between agents, as will be clearer  in Algorithm 2.}
Based on the above discussion, the implementation of the relaxed ADMM method with the $\ell_1$-norm regularization is elaborated as follows:
{\renewcommand{\theenumi}{S\arabic{enumi}}
\begin{enumerate}
	\item Update primal variables: 
\begin{subequations}\label{primal}
	\begin{alignat}{2}
	&{ \{\mathbf{X}^r(t+1)\}, \{\mathbf{Z}^r(t+1)\}}=\min_{( \{\mathbf{X}^r\}, \{\mathbf{Z}^r\})}  \mathcal{L}^{r,1}_{admm}=\notag\\
	& C^r(P_{G}^r)  + \sum_{l\in \mathcal{N}^{\delta{(r)}}}\left[\text{Tr}\left((\mathbf{\Gamma}^r_l(t))^T (\mathbf{E}_{r\mapsto l}\mathbf{X}^r\mathbf{E}_{r\mapsto l}^T\right.\right.\notag\\
	&	\left. -\mathbf{X}^l_r) \right) + \text{Tr}\left((\mathbf{\Lambda}^r_l(t))^T (\mathbf{E}_{r\mapsto l}\mathbf{Z}^r\mathbf{E}_{r\mapsto l}^T-\mathbf{Z}^l_r) \right)+ \notag\\
	& \alpha\norm{\textbf{vec}\left(\mathbf{E}_{r\mapsto l}\mathbf{X}^r\mathbf{E}_{r\mapsto l}^T-\mathbf{X}^l_r\right) }_1+\notag\\
	& \left.\alpha\norm{\textbf{vec}\left(\mathbf{E}_{r\mapsto l}\mathbf{Z}^r\mathbf{E}_{r\mapsto l}^T-\mathbf{Z}^l_r\right) }_1\right]\\
	&\text{s.t.} \{\mathbf{X}^r, \mathbf{Z}^r,  P_G^r, Q_G^r\} \in \mathcal{F}_r^{\mathbb{R}},\\
	& \begin{bmatrix} \mathbf{X}^r & -\mathbf{Z}^r \\ \mathbf{Z}^r & \mathbf{X}^r \end{bmatrix}\succeq 0, 
	\end{alignat}
\end{subequations}
where \eqref{primal} is an approximation to \eqref{Lagrangian} by replacing the  Frobenius norm by the $\ell_1$-norm regularization and $\textbf{vec}(\cdot)$ denotes the operator that reshapes a matrix to a vector.  In addition, $\alpha$ is a fixed positive weighting factor, which is tuned to approach the global optimal solution to (P3).
	\item Update dual variables:

\begin{equation}
\begin{split}\label{dual1}
	\{\mathbf{\Gamma}^r_l(t+1)\}=\{\mathbf{\Gamma}^r_l(t)\}+\rho_{rl} \circ((\mathbf{X}^r_l)(t+1) - \mathbf{X}^l_r)
\end{split}
\end{equation}
\begin{equation}
\begin{split}\label{dual2}
	\{\mathbf{\Lambda}^r_l(t+1)\}=\{\mathbf{\Lambda}^r_l(t)\}+\kappa_{rl}\circ ((\mathbf{Z}^r_l)(t+1) - \mathbf{Z}^l_r)
\end{split}
\end{equation}

\end{enumerate} 
}
In Step (S1), the per-area matrices $\mathbf{X}^r(t+1),\mathbf{Z}^r(t+1)$ are obtained by minimizing \eqref{primal} with $\{\mathbf{\Gamma}^r_l\}$ and $\{\mathbf{\Lambda}^r_l\}$ fixed to their previous iteration values. Likewise, the dual variables $\{\mathbf{\Gamma}^r_l\}$ and $\{\mathbf{\Lambda}^r_l\}$ are updated by \eqref{dual1} and \eqref{dual2} by fixing $\mathbf{X}^r(t+1),\mathbf{Z}^r(t+1)$  to their up-to-date values. In particular, the initial values of $\{\mathbf{\Gamma}^r_l(0)\}$ and $\{\mathbf{\Lambda}^r_l(0)\}$ are zero matrices. Therefore, $\{\mathbf{\Gamma}^r_l(t)\}$ and $\{\mathbf{\Lambda}^r_l(t)\}$, $\forall t, r, l$ are  symmetric  and skew-symmetric matrices, respectively.
  Notice that  $t$ indexes the outer iterations to conduct (S1) and (S2) steps  alternately.


We are ready to write the subgradient of $\mathcal{L}^r$ with respect to $\mathbf{X}^r$ and $\mathbf{Z}^r$. 
\begin{equation}
\begin{split}\label{nabla_x}
&\partial\mathcal{L}^{r,1}_{admm}/\partial\mathbf{X}^r=\sum_{i=1}^{\abs{\mathcal{R}_r}}[2a_i (\mathbf{G}^r_i)^T(\text{Tr}\{(\mathbf{G}^r_i\mathbf{X}^r-\mathbf{B}^r_i\mathbf{Z}^r)\}+P_{D_i})\\
&+b_i (\mathbf{G}^r_i)^T]+\sum_{l\in \mathcal{N}^{\delta{(r)}}} \left[  \mathbf{E}_{r\mapsto l}^T (\mathbf{\Gamma}^r_l+ \alpha{(sign(\mathbf{X}^r_l-\mathbf{X}^l_r))}\mathbf{E}_{r\mapsto l}\right],
\end{split}
\end{equation}
\begin{equation}
\begin{split}\label{nabla_z}
&\partial\mathcal{L}^{r,1}_{admm}/\partial\mathbf{Z}^r=\sum_{i=1}^{\abs{\mathcal{R}_r}}[2a_i (-\mathbf{B}^r_i)^T(\text{Tr}\{(\mathbf{G}^r_i\mathbf{X}^r-\mathbf{B}^r_i\mathbf{Z}^r)\}+P_{D_i})\\
&+b_i (-\mathbf{B}^r_i)^T]+\sum_{l\in \mathcal{N}^{\delta{(r)}}} \left[ \mathbf{E}_{r\mapsto l}^T (\mathbf{\Lambda}^r_l+ \alpha{( sign(\mathbf{Z}^r_l-\mathbf{Z}^l_r))}\mathbf{E}_{r\mapsto l}\right].
\end{split}
\end{equation}
Here,   $a_i = b_i = c_i = 0$ if $i \in 
\mathcal{R}_r / \mathcal{R}^g_r$. $sign(\mathbf{X})$ denotes the sign signals of of each element in $\mathbf{X}$. Define
\begin{equation}
\begin{split}
\mathcal{O}_{rl}	=\{(i,i), (i,j), (j,i), (j, j) \},
\end{split}
\end{equation}
 where line $ij$ connects region $r$ and region $l$. For example, $\mathbf{X}^r_{l, o}, \forall o\in \mathcal{O}_{rl}$ is a element of $\mathbf{X}^r_l$. 
In particular, the subgradients of $\abs{\mathbf{X}^r_{l,o}-{\mathbf{X}^l_{r,o}}}$ and $\abs{\mathbf{Z}^r_{l,o}-{\mathbf{Z}^l_{r,o}}}$ are
\begin{equation}
\begin{split}\label{subgradient_x}
&\partial \abs{\mathbf{X}^r_{l,o}-{\mathbf{X}^l_{r,o}}}/\partial\mathbf{X}^r_{l,o}=\left\{
\begin{aligned}
&	1,   & \mathbf{X}^r_{l,o}>  {\mathbf{X}^l_{r,o}}\\
&	-1, & \mathbf{X}^r_{l,o} <  {\mathbf{X}^l_{r,o}}\\
& [-1, 1], & \mathbf{X}^r_{l,o} =  {\mathbf{X}^l_{r,o}}
\end{aligned}\right.
,
\end{split}
\end{equation}
\begin{equation}
\begin{split}\label{subgradient_z}
&\partial \abs{\mathbf{Z}^r_{l,o}-{\mathbf{Z}^l_{r,o}}}/\partial\mathbf{Z}^r_{l,o}=\left\{
\begin{aligned}
&	1,   & \mathbf{Z}^r_{l,o}>  {\mathbf{Z}^l_{r,o}}\\
&	-1, & \mathbf{Z}^r_{l,o} <  {\mathbf{Z}^l_{r,o}}\\
& [-1, 1], & \mathbf{Z}^r_{l,o} =  {\mathbf{Z}^l_{r,o}}
\end{aligned}\right.
,
\end{split}
\end{equation}
With the above subgradient, we utilize the standard subgradient method to iterate toward the optimal solution \cite{boyd2004convex}. We write the iterative subgradient method to solve \eqref{primal} as
\begin{equation}\label{inner_up}
\begin{split}
     \begin{bmatrix} \mathbf{X}^r(k+1)   \\ \mathbf{Z}^r(k+1)  \end{bmatrix} = \Phi\left(\begin{bmatrix} \mathbf{X}^r(k)  \\ \mathbf{Z}^r(k)  \end{bmatrix}, \begin{bmatrix} (\partial\mathcal{L}^{r,1}_{admm}/\partial\mathbf{X}^r)(k)  \\ (\partial\mathcal{L}^{r,1}_{admm}/\partial\mathbf{Z}^r)(k)   \end{bmatrix}\right),
\end{split}
\end{equation}
where   $\Phi(\cdot)$ denotes the iteration function that will be elaborated in \eqref{update1} and \eqref{update2}. Note that    $k$ indexes the inner iterations to solve \eqref{primal}.
\begin{proposition}
 Problem \eqref{primal} can be  formulated as the standard form:
\begin{subequations}\label{SDP_prop1}
	\begin{alignat}{2}
    \text{Primal: }&\argmin \text{Tr}(A_0(\bm{\partial}))^T X),\\
    & \text{Tr}(A_m^T X)=b_m, \forall m=\{1, \cdots, M\}  \\
     & X\succeq 0.
	\end{alignat}
	\end{subequations}
	where  $A_0(\bm{\partial})$  is relevant to the message communications of the subgradients   and $M$ is the number of constraints. In addition, $A_m\in \mathbb{R}^{D\times D}$, $X\in \mathbb{R}^{D\times D}$ and $b_m \in \mathbb{R}$ will be constructed in the Appendix, where $D$ is the dimension of $X$.
	For conciseness, we omit the notation $\bm{\partial}$ in the following. The dual problem of \eqref{SDP_prop1} is 
	\begin{subequations}\label{SDPxx}
	\begin{alignat}{2}
    \text{Dual: }&\max b^Ty,\\
    & A_0- \sum_{m=1}^M y_mA_m\succeq 0,
	\end{alignat}
	\end{subequations}
	where $b=[b_1, \cdots, b_m]^T$ and $y\in \mathbb{R}^m$ is the vector of dual variables.
\end{proposition}
\emph{Proof:} We show the detailed proof in the Appendix.

Referring to \cite{bai2008semidefinite}, the update scheme $\Phi(\cdot)$  for \eqref{SDPxx} is 
\begin{subequations}\label{update1}
	\begin{alignat}{2}
    & B\Delta y =r\\
    & \Delta X = P+\sum_{m=1}^MA_m \Delta y_m\\
    & \Delta \hat{Z} = (X^k)^{-1} (R- Z^k \Delta X)\\
    & \Delta {Z} = (\hat{Z}+\hat{Z}^T)/2
	\end{alignat}
	\end{subequations}
	where
	\begin{subequations}\label{update2}
	\begin{alignat}{2}
    & B_{mn} = \text{Tr}([(X^k)^{-1}A_mZ^k]^T A_n), \forall m,n=1, \cdots, M,\\
    & r_m=-d_m+ \text{Tr}(A_m^T [(X^k)^{-1}(R-PZ^k)]), \forall m=1, \cdots, M,\\
    & P= \sum_{m=1}^M A_m y_m^k -A_0-X^k,\\
    & d_m = b_m- \text{Tr}(A_m^T Z^k), \forall m=1, \cdots, M,\\
    & R=\beta\mu^kI-X^kZ^k,\\
    & \mu^k = \text{Tr}((X^k)^T Z^k)/D.
	\end{alignat}
	\end{subequations}

%
\begin{figure}[!t]
\centering
\includegraphics[width=3.5in]{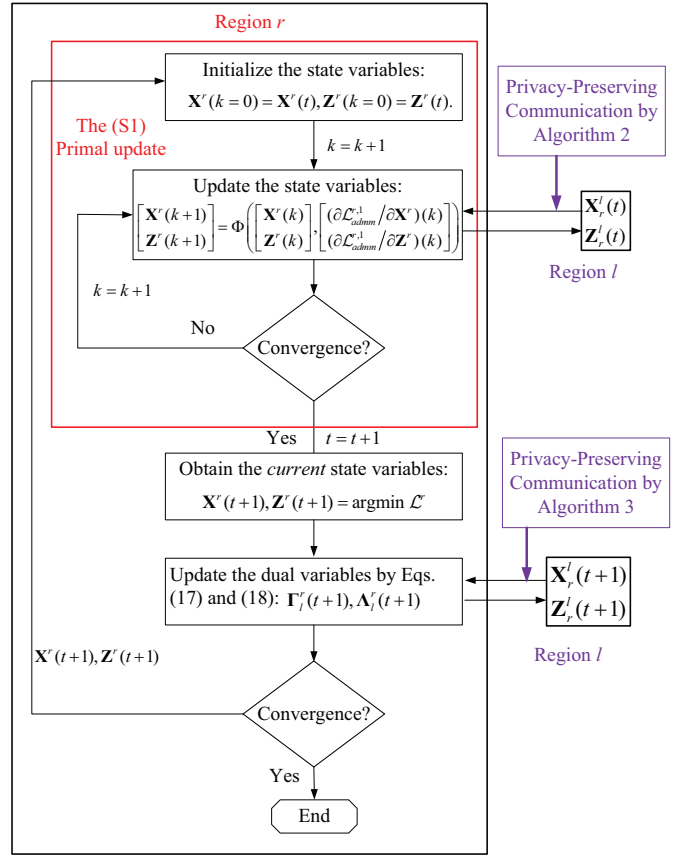}
\caption{The flowchart of the proposed algorithm.}\label{fig02}
\vspace{-0.4cm}
\end{figure}

The above  algorithm cannot  protect the privacy of multiple agents when the messages are exchanged and disclosed explicitly among neighboring agents. To avoid privacy leakage, we propose a privacy preserving distributed optimization  method, which combines  homomorphic cryptography and distributed optimization in the following subsection.

\subsection{Privacy-Preserving Distributed Algorithm}
In this subsection, we combine Paillier cryptosystem with DOPF to enable privacy preservation in the (S1) and (S2) steps. We illustrate the DOPF based on PHE  in Fig. \ref{fig02}.

First of all, the proposed DOPF based on PHE has two loops, namely the inner  (S1) primal update iterations and outer iterations. In the inner  (S1) primal update, we first initialize the  state variables $\mathbf{X}^r(k=0), \mathbf{Z}^r(k=0)$.
 Then, we update the iterative state variables $\mathbf{X}^r(k+1), \mathbf{Z}^r(k+1)$ according to Eq. \eqref{inner_up}, where the state variables during the inner (S1)  update (i.e., the $k$th iteration)  are infeasible to $\mathcal{F}_r^{\mathbb{R}}$ unless the inner (S1) update converges. In contrast, the state variables  during the outer (S2) dual update (i.e., the $t$th iteration) are   always feasible to $\mathcal{F}_r^{\mathbb{R}}$.

  In particular,  Eq. \eqref{inner_up} requires agent $r$ to  successively compare its iterative states (i.e., $\mathbf{X}^r_l(k), \mathbf{Z}^r_l(k)$) from agent $l$'s previous intermediate states (i.e., $\mathbf{X}^l_r(t), \mathbf{Z}^l_r(t)$) in each inner iteration $k$. In this process,  the communication between agent $r$ and agent $l$ can expose its sensitive information to each other or eavesdroppers. Therefore, we propose   Algorithm \ref{alg:A} to enable the privacy-preserving communication in the (S1) primal update.

After the inner  (S1) primal update converges, agent $r$ obtains the next intermediate states $\mathbf{X}^r(t+1), \mathbf{Z}^r(t+1)$.  Then, agent $r$ updates the dual variables  $\{\mathbf{\Gamma}^r_l(t+1)\}$ and $\{\mathbf{\Lambda}^r_l(t+1)\}$ with the help of the communication from agent $l$ about its current intermediate states, i.e., $\mathbf{X}^l_r(t+1), \mathbf{Z}^l_r(t+1)$. However, this process is likely to leak the agents' intermediate states. To address this problem,  we propose Algorithm \ref{alg:B} to facilitate the privacy preserving communication in the dual update.

   \begin{algorithm} [!htb]\small
       \caption{Privacy-Preserving Message Communication for the  (S1) (Primal) update}
    \label{alg:A} 
      \textbf{Agent} $r$ encrypts $\mathbf{X}^{r}_{l,o}(k)$ and $\mathbf{Z}^{r}_{l,o}(k)$  with its public key $k_{pi}$:
\begin{equation}\label{step1}
\begin{split}
  \mathbf{X}^{r}_{l,o}(k) \mapsto \mathcal{E}(\mathbf{X}^{r}_{l,o}(k)), \forall o\in \mathcal{O}_{rl},\\
 \mathbf{Z}^{r}_{l,o}(k) \mapsto \mathcal{E}(\mathbf{Z}^{r}_{l,o}(k)), \forall o\in \mathcal{O}_{rl},   	
\end{split}
\end{equation}
where the superscript $r$ denotes encryption using the public key of  \textbf{Agent} $r$\;
\textbf{Agent} $r$ sends $\mathcal{E}(\mathbf{X}^{r}_{l,o}(k))$ and $\mathcal{E}(\mathbf{Z}^{r}_{l,o}(k))$, and its public key $k_{pi}$ to neighboring \textbf{Agent} $l$\;
\textbf{Agent} $l$ encrypts $\mathbf{X}^{l}_{r,o}(t)$ and $\mathbf{Z}^{l}_{r,o}(t)$ with its public key $k_{pi}$:
\begin{equation}\label{step3}
\begin{split}
      	\mathbf{X}^{l}_{r,o}(t) \mapsto \mathcal{E}(-\mathbf{X}^{l}_{r,o}(t)),\\
      	\mathbf{Z}^{l}_{r,o}(t) \mapsto \mathcal{E}(-\mathbf{Z}^{l}_{r,o}(t)),
\end{split}
\end{equation}
where $\forall o\in \mathcal{O}_{rl}$\;
 \textbf{Agent} $l$  computes the difference directly in ciphertext: 
\begin{equation}\label{step4}
\begin{split}
&\mathcal{E}(\mathbf{X}^{r}_{l,o}(k) -\mathbf{X}^l_{r,o}(t))=\mathcal{E}(\mathbf{X}^{r}_{l,o}(k))\cdot \mathcal{E}(-\mathbf{X}^{l}_{r,o}(t)),\\
&\mathcal{E}(\mathbf{Z}^{r}_{l,o}(k) -\mathbf{Z}^l_{r,o}(t))=\mathcal{E}(\mathbf{Z}^{r}_{l,o}(k))\cdot \mathcal{E}(-\mathbf{Z}^{l}_{r,o}(t)),
\end{split}
\end{equation}
since Paillier system is  additively homomorphic\;
 \textbf{Agent} $l$  computes the $c_{l\mapsto r}(k)$ and $d_{l\mapsto r}(k)$-weighted difference  in ciphertext: 
\begin{equation}\label{step5}
\begin{split}
&\mathcal{E}(c_{l\mapsto r}(k)(\mathbf{X}^{r}_{l,o}(k) -\mathbf{X}^{l}_{r,o}(t)))= \\
&(\mathcal{E}(\mathbf{X}^{r}_{l,o}(k) -\mathbf{X}^{l}_{r,o}(t)))^{c_{l\mapsto r}(k)},\\
&\mathcal{E}(d_{l\mapsto r}(k)(\mathbf{Z}^{r}_{l,o}(k) -\mathbf{Z}^{l}_{r,o}(t)))= \\
&(\mathcal{E}(\mathbf{Z}^{r}_{l,o}(k) -\mathbf{Z}^{l}_{r,o}(t)))^{d_{l\mapsto r}(k)},
\end{split}
\end{equation}
and sends $\mathcal{E}(c_{l\mapsto r}(k)(\mathbf{X}^{r}_{l,o}(k) -\mathbf{X}^{l}_{r,o}(t)))$ and $\mathcal{E}(d_{l\mapsto r}(k)(\mathbf{Z}^{r}_{l,o}(k) -\mathbf{Z}^{l}_{r,o}(t)))$ back to  \textbf{{System Operator}}\;
 \textbf{Agent} $r$ sends its private key $k^a_{si}$ to \textbf{{System Operator}}\;
 \textbf{{System Operator}} decrypts the message received from  \textbf{Agent} $l$ with its private key $k^a_{si}$  to get $c_{l\mapsto r}(k)(\mathbf{X}^{r}_{l,o}(k) -\mathbf{X}^{l}_{r,o}(t))$ and $d_{l\mapsto r}(k)(\mathbf{Z}^{r}_{l,o}(k) -\mathbf{Z}^{l}_{r,o}(t))$\; 
  \textbf{{System Operator}} obtains the sign of the two messages, i.e., $\text{sign}(c_{l\mapsto r}(k)(\mathbf{X}^{r}_{l,o}(k) -\mathbf{X}^{l}_{r,o}(t)))$ and $\text{sign}(d_{l\mapsto r}(k)(\mathbf{Z}^{r}_{l,o}(k) -\mathbf{Z}^{l}_{r,o}(t)))$, and sends the sign messages to  \textbf{Agent} $r$\;
   When the two messages $c_{l\mapsto r}(k)(\mathbf{X}^{r}_{l,o}(k) -\mathbf{X}^{l}_{r,o}(t))$ and $d_{l\mapsto r}(k)(\mathbf{Z}^{r}_{l,o}(k) -\mathbf{Z}^{l}_{r,o}(t))$ $\forall r, l$ are zeros,  \textbf{{System Operator}}  informs all agents that the algorithm converges.  As a result, primal intermediate states at $t+1$th iteration  are obtained by \eqref{primal}\;
    \end{algorithm}
   \begin{algorithm} [!htb]\small
       \caption{Privacy-Preserving Message Communication for the (S2) (Dual) update}
    \label{alg:B} 
      \textbf{Agent} $r$ encrypts $-\mathbf{X}^{r}_{l,o}(t)$ and $-\mathbf{Z}^{r}_{l,o}(t)$  with its public key $k_{pi}$, similar to  \eqref{step1}\;
\textbf{Agent} $r$ sends $\mathcal{E}(-\mathbf{X}^{r}_{l,o}(t))$ and $\mathcal{E}(-\mathbf{Z}^{r}_{l,o}(t))$, and its public key $k_{pi}$ to neighboring \textbf{Agent} $l$\;
      \textbf{Agent} $l$ encrypts $\mathbf{X}^{l}_{r,o}(t)$ and $\mathbf{Z}^{l}_{r,o}(t)$ with its public key $k_{pi}$, similar to \eqref{step3}\;
 \textbf{Agent} $l$  computes the difference, i.e., $\mathcal{E}(\mathbf{X}^{l}_{r,o}(t) -\mathbf{X}^{r}_{l,o}(t))$ and $\mathcal{E}(\mathbf{Z}^{l}_{r,o}(t) -\mathbf{Z}^{r}_{l,o}(t))$, directly in ciphertext, similar to \eqref{step4}\;
 \textbf{Agent} $l$  computes the $a_{l\mapsto r}(t)$ and $b_{l\mapsto r}(t)$-weighted difference, i.e., $\mathcal{E}(a_{l\mapsto r}(t)(\mathbf{X}^l_{r,o}(t) -\mathbf{X}^r_{l,o}(t)))$ and $\mathcal{E}(b_{l\mapsto r}(t)(\mathbf{Z}^l_{r,o}(t) -\mathbf{Z}^r_{l,o}(t)))$  in ciphertext, similar to \eqref{step5}\;
 \textbf{Agent} $l$ sends $\mathcal{E}(a_{l\mapsto r}(t)(\mathbf{X}^l_{r,o}(t) -\mathbf{X}^r_{l,o}(t)))$ and $\mathcal{E}(b_{l\mapsto r}(t)(\mathbf{Z}^l_{r,o}(t) -\mathbf{Z}^r_{l,o}(t)))$ back to  \textbf{Agent} $r$\;
\textbf{Agent} $r$ decrypts the message received from \textbf{Agent} $l$  with its private key $k_{si}$ and multiplies the result with $a_{r\mapsto l}(t)$ and $b_{r\mapsto l}(t)$ to get $\rho_{rl}(t)( \mathbf{X}^r_{l,o}(t)- \mathbf{X}^l_{r,o}(t))$ and $\kappa_{rl}(t)(\mathbf{X}^r_{l,o}(t)- \mathbf{X}^l_{r,o}(t) )$:
\begin{equation}
\begin{split}	
\rho_{rl}(t)=a_{l\mapsto r}(t)\cdot a_{r\mapsto l}(t),\\
\kappa_{rl}(t)=b_{l\mapsto r}(t)\cdot b_{r\mapsto l}(t).
\end{split}
\end{equation}
Computing \eqref{dual1} and \eqref{dual2} to update dual variables in iteration $t$\;

   \end{algorithm}

In the following, we introduce the privacy-preserving communications between participating agents.
In particular, for the (S1) inner update,  Eqs. \eqref{nabla_x} and \eqref{nabla_z} require the exchanged messages, i.e., $sign(\mathbf{X}^r_l-\mathbf{X}^l_r)$ and $sign(\mathbf{Z}^r_l-\mathbf{Z}^l_r)$. 
In Step (S2), we have $(\rho_{rl} \circ (\mathbf{X}^r_l-\mathbf{X}^l_r))$ and $(\kappa_{rl} \circ (\mathbf{Z}^r_l-\mathbf{Z}^l_r))$. They require message communications among regions.  Noticeably, Paillier encryption cannot be performed on matrices directly. Therefore, each element of the matrix, e.g.,  $\mathbf{X}^r_l$ and $\mathbf{Z}^r_l$, are encrypted separately. 
We summarize the Algorithms \ref{alg:A} and  \ref{alg:B} as follows.
\begin{itemize}
	\item For the  (S1) update, agent $l$ generates two random positive scalers, i.e., $c_{l\mapsto r}(k)$ and $d_{l\mapsto r}(k)$, to multiply the  differences of the two messages in the $k$th iteration. Then, the encrypt weighted differences of the two messages  (in ciphertext) are sent to the {system operator}. The {system operator} sends the sign signals to agent $r$. The  privacy-preserving message communication for the primal update is elaborated in Algorithm \ref{alg:A}. 
	\item For the  (S2) update, we construct $\rho_{rl}(t)$ and $\kappa_{rl}(t)$, $r\neq l$ as the product of two random positive numbers, i.e., $\rho_{rl}(t)=a_{r\mapsto l}(t)\cdot a_{l\mapsto r}(t)$ and $\kappa_{rl}(t)=b_{r\mapsto l}(t)\cdot b_{l\mapsto r}(t)$. In particular, $a_{r\mapsto l}(t)$ and $b_{r\mapsto l}(t)$ are only known to agent $r$. This way, we further propose Algorithm \ref{alg:B} to enable the privacy-preserving message communication in the dual update.
\end{itemize}

 \begin{remark}
	In Step 5 of Algorithm \ref{alg:A}, $c_{l\mapsto r}(k)$ and $d_{l\mapsto r}(k)$ are large random positive integers. The two numbers are only known to agent $l$ and varying in each iteration.
\end{remark}
\begin{remark}
	In Steps 1-2 of Algorithm \ref{alg:B}, agent $r$'s states $\mathbf{X}^r_{l,o}$ and $\mathbf{Z}^r_{l,o}$ are encrypted and will not be revealed to its neighbors.
\end{remark}
\begin{remark}
	In Steps 4-5 of Algorithm \ref{alg:B}, agent $l$'s state  will not be revealed to agent $r$ because the decrypted messages obtained by agent $r$ are $a_{l\mapsto r}(t)(\mathbf{X}^l_{r,o}(t) -\mathbf{X}^r_{l,o}(t))$ and $b_{l\mapsto r}(t)(\mathbf{X}^l_{r,o}(t) -\mathbf{X}^r_{l,o}(t))$, where $a_{l\mapsto r}(t)$ and $b_{l\mapsto r}(t)$ are only known to agent $l$ and varying in each iteration.
\end{remark}
\begin{remark}
	Although Pailler cryptosystem only works for integers, we can take additional steps to convert real values in optimization to integers.  This may lead to quantization errors. A common workaround is to scale the real value before quantization.
\end{remark}
\begin{remark}
	The key to achieve the privacy-preserving message communication is to construct $\rho_{rl}, r\neq l$ (or $\kappa_{rl}$) as the product of two random numbers $a_{r\mapsto l}$ and $a_{l\mapsto r}$ (or $b_{r\mapsto l}$ and $b_{l\mapsto r}$) generated by and only known by corresponding agents. Therefore, the convergence of algorithm  is affected by the random, time-varying $\rho_{rl}$ and $\kappa_{rl}$.
\end{remark}

\section{Privacy  Analysis and Stopping Criterion}
In the section, we  analyze the privacy of agents' immediate states, and  convergence of  Algorithms \ref{alg:A} and \ref{alg:B}. In addition, we also prove that  the private information, including  agents' gradients and the objective functions,  cannot be inferred by honest-but-curious adversaries, external eavesdroppers, and  the {system operator} over time. 

\subsection{Privacy Analysis}
\begin{theorem}\label{them1}
	Assume that all agents follow Algorithm \ref{alg:A}. Then agent $l$'s exact state values $\mathbf{X}^l_r(t)$ and $\mathbf{Z}^l_r(t)$ over the boundaries cannot be inferred by an honest-but-curious agent $r$ and the {system operator} unless $\mathbf{X}^r_l(k)=\mathbf{X}^l_r(t)$ and $\mathbf{Z}^r_l(k)=\mathbf{Z}^l_r(t)$.
\end{theorem}
\emph{Proof:} In Algorithm \ref{alg:A}, we have two potential adversaries, i.e., the honest-but-curious agent $r$ and the {system operator}. 
\begin{itemize}
	\item The honest-but-curious agent $r$  only obtains the sign messages from the agent $l$, i.e., ${sign}(c_{l\mapsto r}(k)(\mathbf{X}^r_{l,o}(k) - \mathbf{X}^l_{r,o}(t)))$ and ${sign}(d_{l\mapsto r}(k)$ $(\mathbf{Z}^r_{l,o}(k)-\mathbf{Z}^l_{r,o}(t)))$. Taking $\mathbf{X}$ for example, the honest-but-curious agent $r$ collects information from $K$ iterations:
	\begin{subequations}\label{theorem1eq}
	\begin{alignat}{2}
	&\bm{y}_s^0={sign}(c_{l\mapsto r}(0)((\mathbf{X}^r_{l,o}(0) - \mathbf{X}^l_{r,o}(t))),\notag\\
	&\bm{y}_s^1={sign}(c_{l\mapsto r}(1)((\mathbf{X}^r_{l,o}(1) - \mathbf{X}^l_{r,o}(t))),\notag\\
	&	...\notag\\
	&\bm{y}_s^K={sign}(c_{l\mapsto r}(K)((\mathbf{X}^r_{l,o}(K) - \mathbf{X}^l_{r,o}(t))).\notag
\end{alignat}
\end{subequations}
To the honest-but-curious agent $r$, the sign signals   $\bm{y}_s^k$ and $\mathbf{X}^r_{l,o}(k)$ are known, while $\mathbf{X}^l_{r,o}(t)$ and $c_{l\mapsto r}(k)$ are unknown. Therefore, $\mathbf{X}^l_{r,o}(t)$  cannot be inferred by the honest-but-curious agent $r$.
	\item The third-party {system operator} only obtains the weighted differences of two messages from the agent $l$, i.e., $(c_{l\mapsto r}(k)(\mathbf{X}^r_{l,o}(k) - \mathbf{X}^l_{r,o}(t)))$ and $(d_{l\mapsto r}(k)$ $(\mathbf{Z}^r_{l,o}(k)-\mathbf{Z}^l_{r,o}(t)))$. Taking $\mathbf{X}$ for example, the third-party {system operator} collects information from $K$ iterations:
	\begin{subequations}\label{theorem2eq}
	\begin{alignat}{2}
	&\bm{y}_t^0=(c_{l\mapsto r}(0)((\mathbf{X}^r_{l,o}(0) - \mathbf{X}^l_{r,o}(t))),\notag\\
	&\bm{y}_t^1=(c_{l\mapsto r}(1)((\mathbf{X}^r_{l,o}(1) - \mathbf{X}^l_{r,o}(t))),\notag\\
	&	...\notag\\
	&\bm{y}_t^K=(c_{l\mapsto r}(K)((\mathbf{X}^r_{l,o}(K) - \mathbf{X}^l_{r,o}(t))).\notag
\end{alignat}
\end{subequations}
To the third-party {system operator}, only the  signals   $\bm{y}_t^k$  are  known, while $\mathbf{X}^r_{l,o}(k)$, $\mathbf{X}^l_{r,o}(t)$ and $c_{l\mapsto r}(k)$ are unknown. Therefore, $\mathbf{X}^r_{l,o}(k), \forall k=1, \dots, K$ and $\mathbf{X}^l_{r,o}(t)$ cannot be inferred by the the third-party {system operator}.

	\end{itemize}
This completes the proof.
\hfill $\blacksquare$

\begin{theorem}\label{them1}
	Assume that all agents follow Algorithm \ref{alg:B}. Then agent $l$'s exact state values $\mathbf{X}^l_r(t)$ and $\mathbf{Z}^l_r(t)$ over the boundaries cannot be inferred by an honest-but-curious agent $r$ unless $\mathbf{X}^r_l(t)=\mathbf{X}^l_r(t)$ and $\mathbf{Z}^r_l(t)=\mathbf{Z}^l_r(t)$.
\end{theorem}
\emph{Proof:} Suppose that an honest-but-curious agent $r$ collects information from $T$ outer iterations to infer the information of  a neighboring agent $l$. From the perspective of adversary agent $r$, the  measurements received by agent $r$ in the outer $t$th iteration  are $\bm{y}_x^t=a_{r\mapsto l}(t)\cdot a_{l\mapsto r}(t)(\mathbf{X}^r_{l,o}(t) - \mathbf{X}^l_{r,o}(t))$ and $\bm{y}_z^t=b_{r\mapsto l}(t)\cdot b_{l\mapsto r}(t)(\mathbf{Z}^r_{l,o}(t) - \mathbf{Z}^l_{r,o}(t))$ $(o\in \mathcal{O}_{rl}, t=0, 1, \cdots, T)$. In this way, the honest-but-curious agent $r$ can establish $(T+1)\times \abs{\mathcal{O}_{rl}}\times 2$ equations. In the following, we consider one element of  $\mathbf{X}^r_l$. 
\begin{equation}
\begin{split}\label{theorem2eq}
	&\bm{y}_x^0=a_{r\mapsto l}(0)\cdot a_{l\mapsto r}(0)((\mathbf{X}^r_{l,o}(0) - \mathbf{X}^l_{r,o}(0)),\\
	&\bm{y}_x^1=a_{r\mapsto l}(1)\cdot a_{l\mapsto r}(1)((\mathbf{X}^r_{l,o}(1) - \mathbf{X}^l_{r,o}(1)),\\
	&	...\\
	&\bm{y}_x^T=a_{r\mapsto l}(T)\cdot a_{l\mapsto r}(T)((\mathbf{X}^r_{l,o}(T) - \mathbf{X}^l_{r,o}(T)).
\end{split}
\end{equation}
To the honest-but-curious agent $r$ in the above system, $\bm{y}_x^t$, $a_{r\mapsto l}(t)$, $\mathbf{X}^r_{l,o}(t)$ ($t=0, 1, \cdots, T$) are known, but  $a_{l\mapsto r}(t)$ and $\mathbf{X}^l_{r,o}(t)$ ($t=0, 1, \cdots, T$) are unknown. Therefore, the  system of Eq. \eqref{theorem2eq} over $T+1$ equations contains $2*(T+1)$ unknown variables. Therefore, the honest-but-curious agent $r$ cannot solve the system of Eq. \eqref{theorem2eq} to infer the exact values of unknowns $a_{l\mapsto r}(t)$ and $\mathbf{X}^l_{r,o}(t)$ ($t=0, 1, \cdots, T$) of agent $l$, except when $\bm{y}_x^t=0$ and $\mathbf{X}^r_{l,o}(t) = \mathbf{X}^l_{r,o}(t)$. 

This completes the proof.
\hfill $\blacksquare$

\begin{theorem}
	Assume that all agents follow Algorithms \ref{alg:A} and \ref{alg:B}. Then exact gradient  of $C^l$ of agent $l$  cannot be inferred by an honest-but-curious agent $r$. 
\end{theorem}
\emph{Proof:} Recall that 
\begin{equation}\small
\begin{split}
&\nabla_{\mathbf{X}^l} C^l_i = [2a_i (\mathbf{G}^l_i)^T(\text{Tr}\{(\mathbf{G}^l_i\mathbf{X}^l-\mathbf{Z}^l_i\mathbf{Z}^l)\}+P_{D_i})+b_i (\mathbf{G}^l_i)^T],\\
&\nabla_{\mathbf{Z}^l} C^l_i = [2a_i (-\mathbf{Z}^l_i)^T(\text{Tr}\{(\mathbf{G}^l_i\mathbf{X}^l-\mathbf{Z}^l_i\mathbf{Z}^l)\}+P_{D_i})-b_i (\mathbf{Z}^l_i)^T].\notag
\end{split}
\end{equation}

If the generator $G_i$ is not on the boundary of region $r$ and region $l$,  the messages $\mathbf{X}^l_r$ and $\mathbf{Z}^l_r$ do not include the  generator $G_i$. Therefore, the above gradient  of $C^l$ of agent $l$ cannot be inferred by an honest-but-curious agent $r$.

Otherwise, Theorem \ref{them1} proves that the messages $\mathbf{X}^l_r$ and $\mathbf{Z}^l_r$ cannot be inferred by an honest-but-curious agent $r$. Therefore, $\nabla_{\mathbf{X}^l} C^l_i$ and $\nabla_{\mathbf{Z}^l} C^l_i$ cannot be inferred.
\hfill $\blacksquare$

\begin{corollary}\label{cor1}
	Assume that all agents follow Algorithm \ref{alg:A}. The agent $l$'s intermediate states, gradients of objective functions, and objective functions cannot be inferred by an external eavesdropper.
\end{corollary}
\emph{Proof:} Since all exchanged messages are encrypted, an external eavesdropper cannot learn anything by intercepting these messages. In addition, the sign signals do not expose any state information to eavesdroppers.
Therefore, it cannot infer any agent's intermediate states, gradients of objective functions, and objective functions.
\hfill $\blacksquare$

\begin{corollary}
	Assume that all agents follow  Algorithm \ref{alg:B}. Both the agent $r$  and  agent $l$ 's intermediate states, gradients of objective functions, and objective functions cannot be inferred by an external eavesdropper and the third party. 
\end{corollary}
\emph{Proof:} Following a similar line to the reasoning of Corollary \ref{cor1} and Theorem \ref{them1}, we can obtain  Corollary 2.
\hfill $\blacksquare$

\subsection{Stopping Criterion}

We define the  residual as follows.
\begin{equation}
	\begin{split}
		\Psi_r=\norm{\begin{bmatrix} \mathbf{X}^r_l  \\ \mathbf{Z}^r_l  \end{bmatrix} - \begin{bmatrix} \mathbf{X}^l_r  \\ \mathbf{Z}^l_r  \end{bmatrix}}_2.
	\end{split}
\end{equation}
The residue gives rise to the following stopping criterion:
\begin{equation}
	\begin{split}
		\Psi_r (t+1)\le \epsilon, \forall r=1, \cdots, R,
	\end{split}
\end{equation}
where $\Psi_r (t+1)$ is the primal residue after $t+1$ iterations. If $\Psi_r (t+1),\forall r$ approach zeros, the feasibility of the primal variables  and the convergence of the dual variables  are both  satisfied. 
\begin{figure}[!htb]
\centering
\includegraphics[width=3.4in]{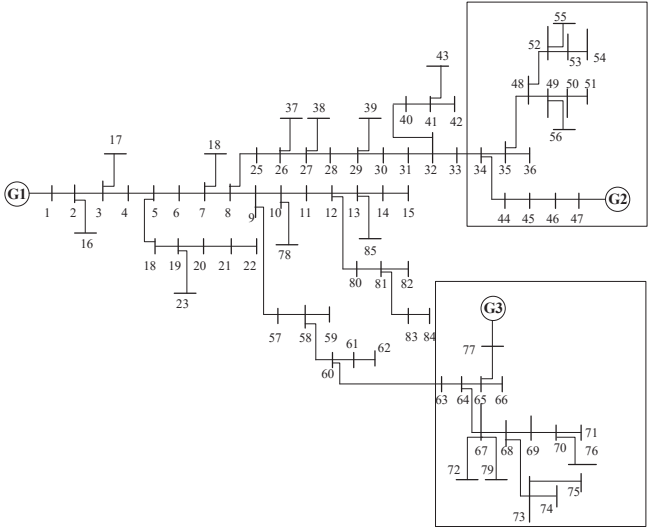}
\caption{The 85-bus tree distribution system with three regions.}\label{tree85}
\vspace{-0.4cm}
\end{figure}
\begin{figure}[!htb]
\centering
\includegraphics[width=3.2in]{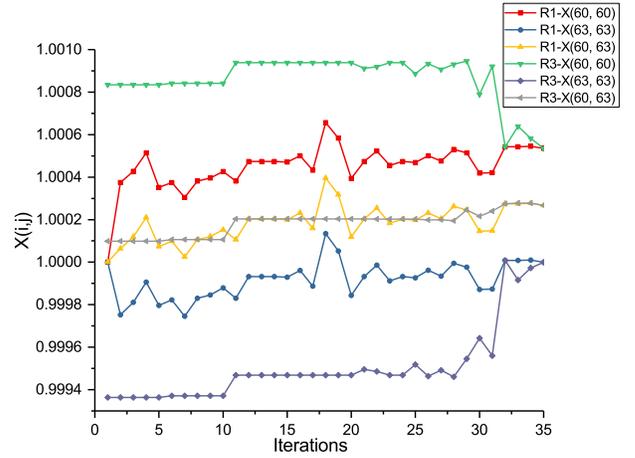}
\caption{The evolution of    $\mathbf{X}^r_{l, o}$.}\label{case1_message}
\vspace{-0.4cm}
\end{figure}
\section{Case Studies}
In this section, we adopt the 85-bus tree distribution system to validate the proposed privacy-preserving distributed OPF algorithm. We partition the    system into three regions, as shown in Fig. \ref{tree85}. 
In particular, two AC generators are added to buses 47 and 77, where fuel cost parameters are $a_i=0.2$, $b_i=2$ and $c_i=2$. Other data for the 85-bus system can be found in MATPOWER's library. We set $\epsilon=10^{-6}$ as the termination criterion of the distributed algorithm.  
\begin{figure}[!htb]
\centering
\includegraphics[width=3.2in]{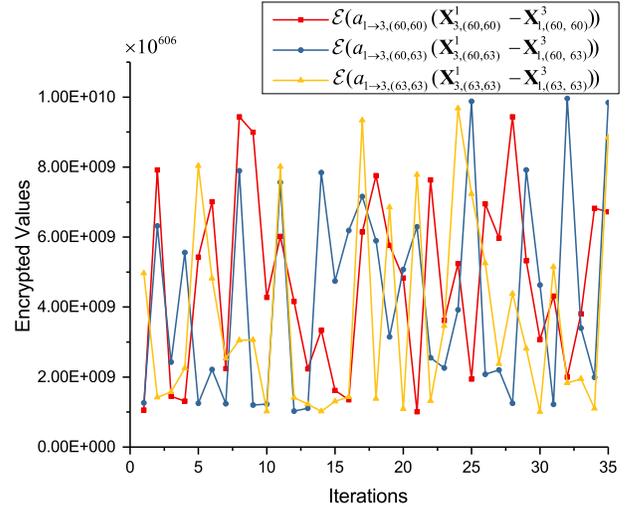}
\caption{The evolution of encrypt weighted differences (in ciphertext).}\label{case1_encrypt_message}
\vspace{-0.4cm}
\end{figure}
\begin{figure}[!htb]
\centering
\includegraphics[width=3.2in]{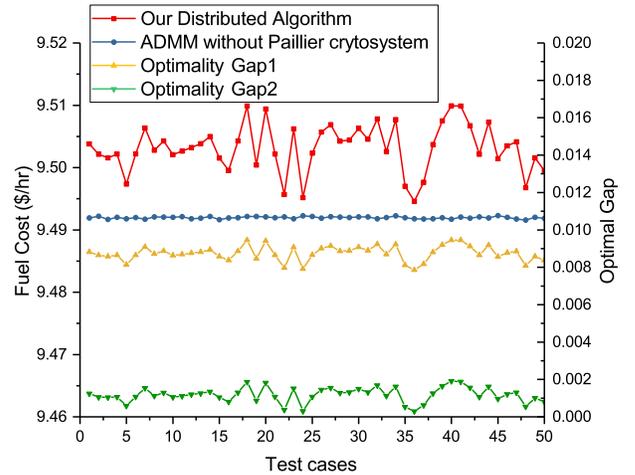}
\caption{The optimal values of the proposed algorithm, the ADMM algorithm, and optimal gaps.}\label{case1_optimal_vs_true}
\vspace{-0.4cm}
\end{figure}
\begin{figure*}[!htb]
\centering
\subfigure[The noise bound $\frac{1}{t^2}$.]{
    \includegraphics[width=0.31\textwidth]{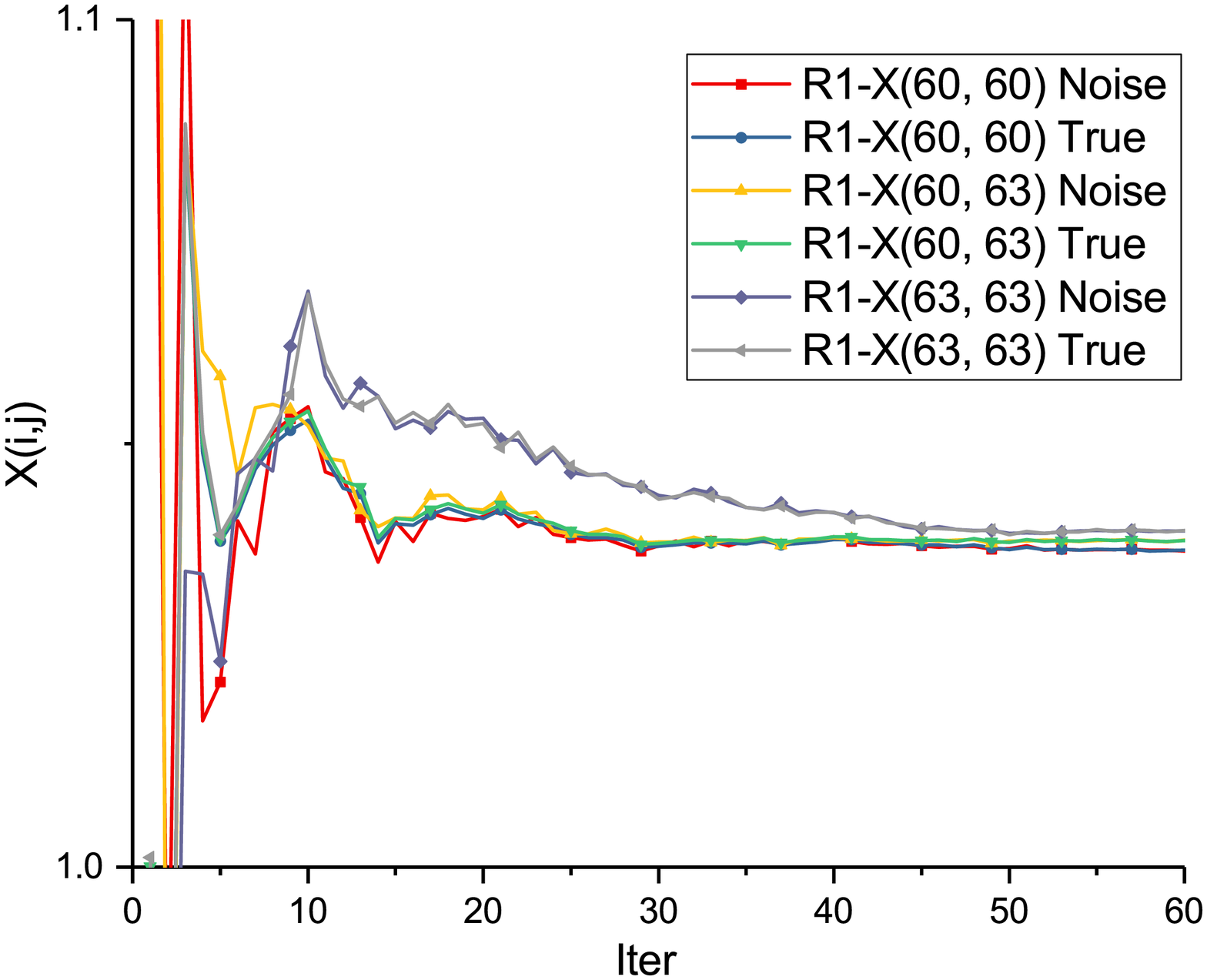}
        \label{fig:case2_message_1_t2}
}
\subfigure[The noise bound $\frac{1}{t^3}$.]{
    \includegraphics[width=0.31\textwidth]{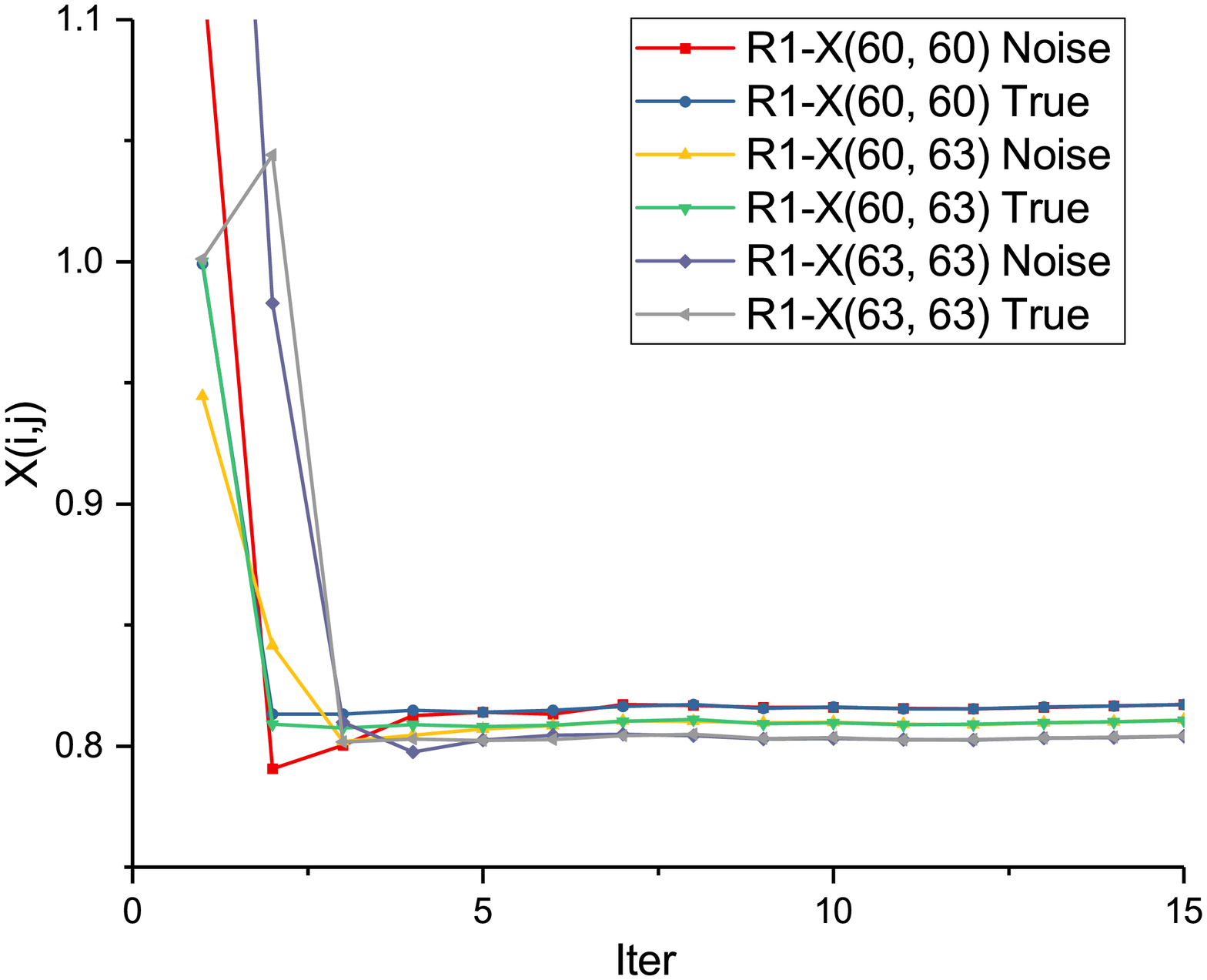}
        \label{fig:case2_message_1_t3}
}
\subfigure[The noise bound $\frac{1}{e^t}$.]{
    \includegraphics[width=0.31\textwidth]{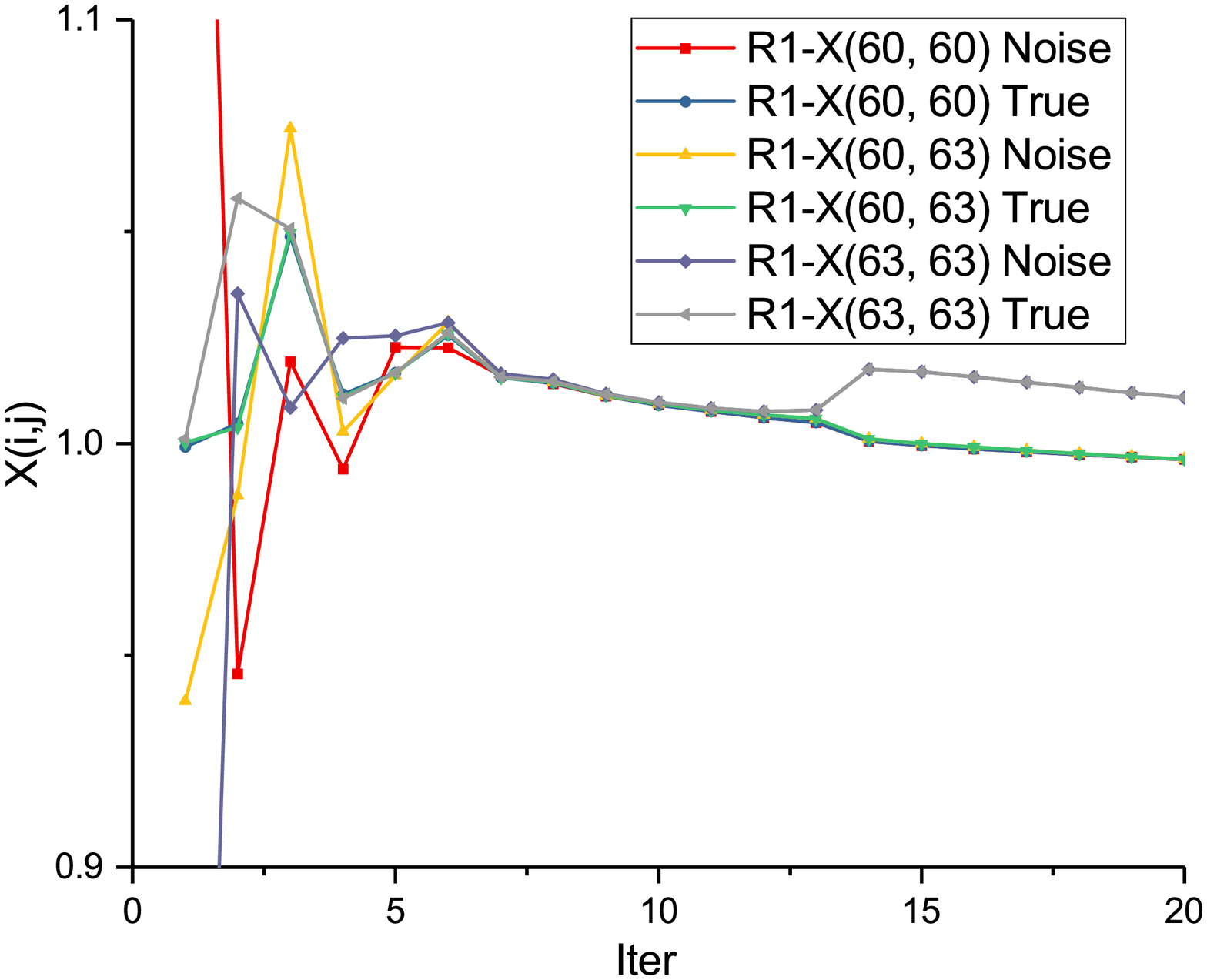}
        \label{fig:case2_message_1_expt}
}
\caption{
The evolution of  $\mathbf{X}^r_{l, o}$  in three noise bounds {of differential privacy methods}.
\label{fig:case2_message}
}
\vspace{-0.4cm}
\end{figure*}

We set $N_{max}=10^{10}$ to convert each element in $\mathbf{X}$ and $\mathbf{Z}$ to a 64-bit integer during intermediate computation. $a_{r\mapsto l}(t)$, $b_{r\mapsto l}(t)$, $c_{r\mapsto l}(k)$ and $d_{r\mapsto l}(k)$ are also scaled to  64-bit integers, respectively. In particular, $a_{r\mapsto l}(t)$, $b_{r\mapsto l}(t)$, $c_{r\mapsto l}(k)$ and $d_{r\mapsto l}(k)$ are  random integers between [100, 200]. $\alpha$ in \eqref{primal} is set to 0.48.  Therefore, $\rho_{rl}$ and $\kappa_{rl}$ vary between [$10^4$, $4\times 10^4$].  The encryption and decryption keys are chosen to be 1028-bit long.  All algorithms are executed on a 64-bit Mac with 2.4 GHz (Turbo Boost up to 5.0GHz) 8-Core Intel Core i9 and a total of 16 GB RAM.

\subsection{Evaluation of Our Approach}
Fig. \ref{case1_message} illustrates the evolution of optimal solution of $\mathbf{X}^r_{l,o}$ in one specific run of Algorithm 3. After 35 iterations, the algorithm converges. Fig. \ref{case1_encrypt_message} visualizes the encrypted weighted differences (in ciphertext) of the states, i.e., $\mathcal{E}(a_{1\mapsto 3, (60, 60)}(\mathbf{X}_{3, (60, 60)}^1 - \mathbf{X}_{1, (60, 60)}^3))$, $\mathcal{E}(a_{1\mapsto 3, (60, 63)}$ $(\mathbf{X}_{3, (60, 63)}^1 - \mathbf{X}_{1, (60, 63)}^3))$ and $\mathcal{E}(a_{1\mapsto 3, (63, 63)}(\mathbf{X}_{3, (63, 63)}^1 - \mathbf{X}_{1, (63, 63)}^3))$. Although the states of all agents converge after 36 iterations, the encrypted weighted differences (in ciphertext) are random to an outside eavesdropper. In addition, it takes about 1ms for each agent to finish the encryption/decryption process  at each iteration. Such a communication delay can be omitted compared with the time cost of each iteration.

\begin{figure}[!htb]
\centering
\includegraphics[width=3.2in]{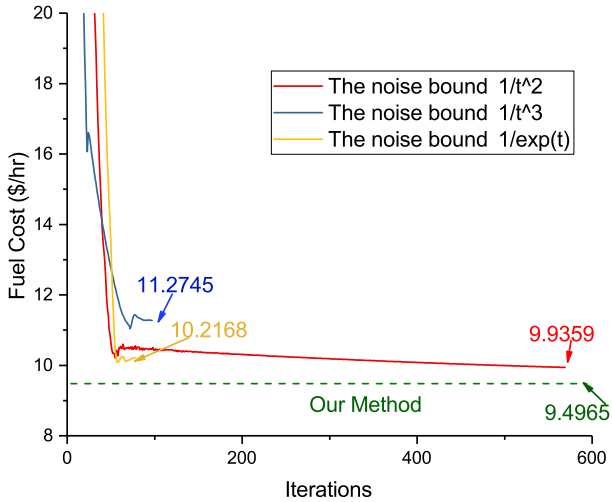}
\caption{The evolution of fuel costs under three  noise bounds vs the optimal value achieved by our algorithm.}\label{case2_noise_value}
\vspace{-0.4cm}
\end{figure}
\begin{figure}[!htb]
\centering
\includegraphics[width=3.2in]{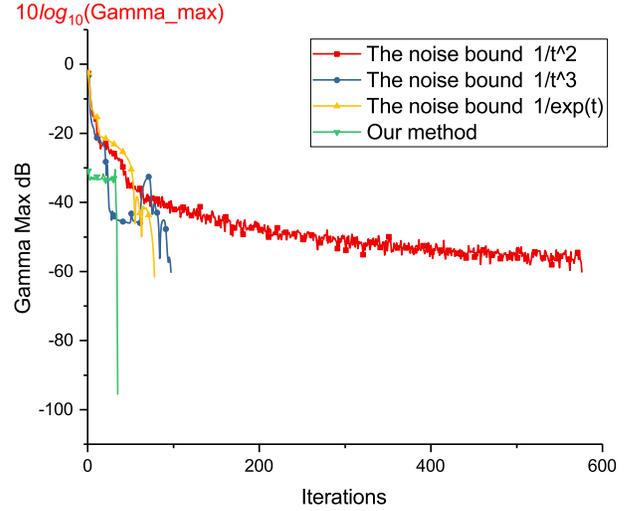}
\caption{Convergence of the proposed algorithm VS differential-privacy methods.}\label{case3_gamma}
\vspace{-0.4cm}
\end{figure}

In Fig. \ref{case1_optimal_vs_true},  we show that the optimality gap of the proposed algorithm is actually  small. {We first define the optimal solution $\bm{x}^{*}$ as the solution to (P1) without the rank-1 constraint, which becomes a convex SDP that can be solved in a centralized manner. Due to the tree structure of the network and some other technical conditions being fulfilled, this SDP relaxation   is exact.} The global optimal solution is solved by the convex optimization solver, i.e., 9.42 \$/hr.    We also define the  distributed solution $\bm{x}^{*}_d$ as the solution to (P4) obtained by ADMM without Paillier cryptosystem. The  solutions obtained by  Algorithms 2\&3 are denoted by  $\bm{x}^{*}_p$.
    Let us denote $z^*$ as the optimal value to   problem (P1), and $z^{*}_d$ and $z^{*}_p$ as the near-optimal values corresponding to $\bm{x}^{*}_d$ and $\bm{x}^{*}_p$, respectively. 
Here, the percentage optimality gap  is calculated as gap$_1$ =100 $\times$ $(z^{*}_p-z^*) /z^{*}_p$ \%. Compared with the ADMM without Paillier cryptosystem, the gap  is calculated as gap$_2$ =100 $\times$ $(z^{*}_p-z^*_d) /z^{*}_p$ \%.
%
%
 Fig. \ref{case1_optimal_vs_true} shows that both gap$_1$ and gap$_2$ are very small, i.e., on average $0.87540 \%$ and $0.11815 \%$  for gap$_1$ and gap$_2$, respectively. This indicates that the solutions by the proposed algorithms are very close to the global optimum.

\subsection{Comparison with the Differential-Privacy Algorithm}
We then compare our approach with the differential-privacy distributed optimization algorithm. In the differential-privacy distributed optimization algorithm, the messages are added with random noises. For example, a random noise $n^{r, x, t}_{l,o}$ is added into $\mathbf{X}^{r, t}_{l,o}$, denoted by $\mathbf{\tilde{X}}^{r, t}_{l,o}= \mathbf{X}^{r, t}_{l,o} + n^{r, x, t}_{l,o}$. Similarly,  $n^{r, z, t}_{l,o}$ is added into $\mathbf{Z}^{r, t}_{l,o}$, denoted by $\mathbf{\tilde{Z}}^{r, t}_{l,o}= \mathbf{Z}^{r, t}_{l,o} + n^{r, z, t}_{l,o}$. In \cite{mao2020privacy}, both $n^{r, x, t}_{l,o}$ and $n^{r, z, t}_{l,o}$ are bounded in $[-\hat{n}^{r, t}_{l,o}, \hat{n}^{r, t}_{l,o}]$. The bound $\hat{n}^{r, t}_{l,o}\in \mathbb{R}^+$ satisfies the following conditions:
 \begin{equation}
\begin{split}	
	\sum_{t=0}^\infty \hat{n}^{r, t}_{l,o}<\infty, \sum_{t=0}^\infty (\hat{n}^{r, t}_{l,o})^2<\infty.
\end{split}
 \end{equation}
Therefore, we choose three kinds of harmonic series, i.e., $\hat{n}^{r, t}_{l,o}= \frac{\hat{n}^{r, 0}_{l,o}}{t^2}$, $\hat{n}^{r, t}_{l,o}=\frac{\hat{n}^{r, 0}_{l,o}}{t^3}$ and $\hat{n}^{r, t}_{l,o}=\hat{n}^{r, 0}_{l,o} {e^{-t}}$ for comparison. Fig. \ref{fig:case2_message} visualizes the true messages and the messages with noises. In particular, Fig. \ref{fig:case2_message_1_t2} shows that  the differences between the true messages and the messages with noises are obvious before 20 iterations, while messages with noises are very close to the true messages after 20 iterations until convergence. Therefore, both  honest-but-curious agents and eavesdroppers can infer the true messages easily. Similarly, other harmonic series are likely to disclose the true messages, as shown in Fig. \ref{fig:case2_message_1_t3} and Fig. \ref{fig:case2_message_1_expt}.

 Fig. \ref{case2_noise_value} shows that the case with the noise bound $\frac{\hat{n}^{r, 0}_{l,o}}{t^2}$ converges to 9.9359 \$/hr, the case with the noise bound $\frac{\hat{n}^{r, 0}_{l,o}}{t^3}$ converges to 11.2745 \$/hr and the case with the noise bound $\hat{n}^{r, 0}_{l,o} {e^{-t}}$ converges to 10.2168 \$/hr. In comparison, our algorithm achieves 9.4965 \$/hr of fuel costs, which is very close to the global optimal (i.e., 9.42 \$/hr) with a gap of $0.80556\%$. This is because the added noises disturb the shared messages and inevitably compromise the accuracy of optimization results, leading to a trade-off between privacy and accuracy.




\subsection{Convergence of Our Approach}
\begin{figure}[!htb]
\centering
\includegraphics[width=3.2in]{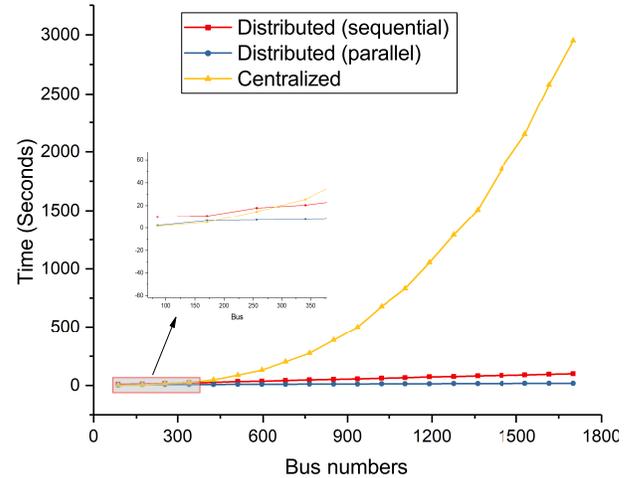}
\caption{Time comparison.}\label{time_comp}
\vspace{-0.4cm}
\end{figure}

The maximal residue of Algorithm \ref{alg:A} and \ref{alg:B} at the $t$th iteration is defined as $\Psi^t_{max}=\max_{r=1, \cdots, R}\Psi_r^t$. Fig. \ref{case3_gamma} shows that  the cases with  noise bounds $\hat{n}^{r, t}_{l,o}= \frac{\hat{n}^{r, 0}_{l,o}}{t^2}$, $\hat{n}^{r, t}_{l,o}=\frac{\hat{n}^{r, 0}_{l,o}}{t^3}$ and $\hat{n}^{r, t}_{l,o}=\hat{n}^{r, 0}_{l,o} {e^{-t}}$ take 79, 97 and 576 iterations to converge, while the proposed algorithm converges within 35 iterations.
 It indicates that the proposed algorithm converges much faster than the differential-privacy distributed algorithm. This is because that the noise-polluted messages  cannot reach a consensus before the noise attenuates to zero.
{
\subsection{Comparison with Other Methods}
We compare the proposed method with other competing alternatives, namely the distributed SDP method \cite{dall2013distributed}, the ADMM method for OPF problems  \cite{erseghe2014distributed}, and the relaxed ADMM method \cite{bastianello2020tac}. The simulation shows that compared with  Ref. \cite{dall2013distributed},  the proposed algorithm (PPOPF) only sacrifices a little optimality with only $0.0295\%$ gap to achieve the privacy preservation guarantee. Besides, the proposed algorithm can converge much faster than other benchmark algorithms. {It is observed from Table \ref{table2} that the proposed PPOPF method can achieve a better such trade-off, e.g., obvious improvement in convergence rate with minor increase of cost. Moreover, compared with differential-privacy methods, the proposed method  obtains the solutions that are closer to the global optimum under a reasonable random penalty range.}

\begin{table}[!h]\footnotesize
\center
\begin{threeparttable}
\caption{PPOPF  VS  other distributed algorithms}
\label{table2}
\centering
\begin{tabular}{c |c c }
\toprule
Methods& Cost & Convergence   \\
\midrule
PPOPF with $\rho_{rl}, \kappa_{rl} \in  [10^4$, $4.00\times 10^4]$ & 9.4946 \$/hr & 35 iterations  \\
PPOPF with $\rho_{rl}, \kappa_{rl} \in  [10^4$, $ 6.25\times 10^4]$ & 9.5029 \$/hr & 30 iterations  \\
PPOPF with $\rho_{rl}, \kappa_{rl} \in  [10^4$, $7.84\times 10^4]$ & 9.5260 \$/hr & 9 iterations  \\
PPOPF with $\rho_{rl}, \kappa_{rl} \in  [10^4$, $9.00\times 10^4]$ & 9.5575 \$/hr & 4 iterations  \\
Distributed SDP  \cite{dall2013distributed}  & 9.4918 \$/hr & 48 iterations    \\
 ADMM method  \cite{erseghe2014distributed}  & 9.4402 \$/hr & 28 iterations  \\
Relaxed ADMM  \cite{bastianello2020tac}      & 9.4917 \$/hr & 48 iterations  \\
\bottomrule
\end{tabular}
\end{threeparttable}
\vspace{-0.3cm}
\end{table}

    $\quad$  First of all, we compare the proposed algorithm with the centralized algorithm in terms of computation time. In Fig. \ref{time_comp}, we show the computational times for test systems of different sizes under three algorithms. In particular, we connect multiple 85-bus systems to construct  large systems with from 85 buses to 1700 buses.
    Here, we only consider the CPU time spent on the SDP solver and assume that communication overheads can be neglected. In Fig. \ref{time_comp},  the sequential distributed algorithm is to sum the CPU times of the subproblems and the parallel distributed algorithm is to consider  the longest CPU time of the subproblems in each iteration. The result shows that the proposed distributed algorithm performs much better than the centralized method.
}     

\section{Conclusion}
This paper proposed a novel privacy-preserving DOPF algorithm based on PHE. The proposed algorithm utilized the ADMM method with the SDP relaxation to solve the OPF problem. For the dual update of the ADMM, we utilized the PHE to encrypt the difference of state variables across neighboring agents. As a result, neither the eavesdroppers nor honest-but-curious agents can infer the exact states. For the primal update, we  relaxed the augmented term of the primal update of ADMM with the $\ell_1$-norm regularization, and then utilized the sign message communications to enable the privacy-preserving primal update. At last, we proved the privacy preservation guarantee of the proposed algorithm. In addition, numerical case studies showed that the  optimality gap of the proposed algorithm is actually small, i.e.,  only $0.8754\%$ between the  solution obtained by the proposed algorithm and the global optimum. Compared with the differential-privacy distributed optimization methods, the proposed method yields a better solution, and converges much faster.

\appendix
In the appendix, we  prove  Proposition 1 and Proposition 2. First of all, let $d_{g_i}$ be an auxiliary variable, which  always equals to one. Obviously, $d_{g_i}^2=1$, $P_{G_i}=P_{G_i}d_{g_i}$ and $Q_{G_i}=Q_{G_i}d_{g_i}$. In the meantime, by introducing  slack variables $u_{g_i}$, $l_{g_i}$, $u_{r_i}$, $l_{r_i}$, $u_{b_i}$ and $l_{b_i}$, inequality constraints \eqref{feasible2} can be transformed into equality constraints as  $u^2_{g_i}\ge 0$, $l^2_{g_i}\ge 0$, $u^2_{r_i}\ge 0$, $l^2_{r_i}\ge 0$, $u^2_{b_i}\ge 0$ and $l^2_{b_i}\ge 0$.
\subsection{Proposition 1}

Problem \eqref{primal} in region $r$ can be equivalently written as:
\begin{subequations}\label{SDP_OLD}
\begin{alignat}{2}
&\argmin&\phantom{xx}&	\sum_{i=1}^{\abs{\mathcal{R}_r^g}}a_iP_{G_i}^2+b_iP_{G_i}+c_i + \sum_{l\in \mathcal{N}^{\delta{(r)}}}\notag\\
&\text{ }&\phantom{xx}&\left[\text{Tr}\left((\mathbf{\Gamma}^r_l)^T (\mathbf{X}^r_l-\mathbf{X}^l_r) \right) + \text{Tr}\left((\mathbf{\Lambda}^r_l)^T (\mathbf{Z}^r_l-\mathbf{Z}^l_r) \right) \right.\notag\\
&\text{ }&\phantom{xx}& \left.+\alpha\norm{\textbf{vec}(\mathbf{X}^r_l-\mathbf{X}^l_r)}_1+\alpha\norm{\textbf{vec}(\mathbf{Z}^r_l-\mathbf{Z}^l_r)}_1\right] \label{obj1_app}\\
&\text{s.t.}&\phantom{xx}& P_{G_i}d_{g_i}-P_{D_i} = \text{Tr}\{\mathbf{Y}_i^r\mathbf{W}^r\},  i\in \mathcal{R}_r^g,\label{SDPconst1}\\
&\text{ }&\phantom{xx}& Q_{G_i}d_{g_i}-Q_{D_i} = \text{Tr}\{\bar{\mathbf{Y}}_i^r \mathbf{W}^r\},i\in \mathcal{R}_r^g,\\
&\text{ }&\phantom{xx}& -P_{D_i} = \text{Tr}\{\mathbf{Y}_i^r\mathbf{W}^r\},  i\in \mathcal{B}_r/\mathcal{R}_r^g,\\
&\text{ }&\phantom{xx}&  -Q_{D_i} = \text{Tr}\{\bar{\mathbf{Y}}_i^r \mathbf{W}^r\},i\in \mathcal{B}_r/\mathcal{R}_r^g,\\
&\text{ }&\phantom{xx}& \text{Tr}\{{\mathbf{M}}_i^r\mathbf{W}^r\}  + u_{bi}^2 = \overline{V_{i}}^2, i\in \mathcal{B}_r, \label{SDPconst2}\\
&\text{ }&\phantom{xx}& \text{Tr}\{{\mathbf{M}}_i^r\mathbf{W}^r\} - l_{bi}^2 = \underline{V_{i}}^2, i\in \mathcal{B}_r,\\
&\text{ }&\phantom{xx}& {P}_{G_i}d_{g_i}+ u_{gi}^2 = \overline{P}_{G_i}, i\in \mathcal{R}_r^g, \label{SDPconst3}\\
&\text{ }&\phantom{xx}& {P}_{G_i}d_{g_i}- l_{gi}^2 = \underline{P}_{G_i}, i\in \mathcal{R}_r^g,\\
&\text{ }&\phantom{xx}& {Q}_{G_i}d_{g_i}+ u_{ri}^2 = \overline{Q}_{G_i}, i\in \mathcal{R}_r^g,\\
&\text{ }&\phantom{xx}& {Q}_{G_i}d_{g_i}- l_{ri}^2 = \underline{Q}_{G_i}, i\in \mathcal{R}_r^g,\\
&\text{ }&\phantom{xx}&d_{g_i}^2=1, i\in \mathcal{R}_r^g,\\
&\text{ }&\phantom{xx}&  \mathbf{W}^r\succeq 0. \label{SDPconst_last1}
\end{alignat}
\end{subequations}

We transform the objective \eqref{obj1_app} as  
\begin{subequations}
\begin{alignat}{2}
&\argmin&\phantom{}&  \sum_{i=1}^{\abs{\mathcal{R}_r^g}}a_iP_{G_i}^2+b_iP_{G_i}+c_i + \sum_{l\in \mathcal{N}^{\delta{(r)}}}\notag\\
&\text{ }&\phantom{}& \left[\text{Tr}\left((\mathbf{\Gamma}^r_l + j\mathbf{\Lambda}^r_l)^H \mathbf{W}^r_l \right)+\alpha\sum_{o\in \mathcal{O}_{rl}}  \right. \notag\\
&\text{ }&\phantom{}& \left. \left(\abs{\mathbf{X}^r_{l,o}-\mathbf{X}^l_{r,o}}+\abs{\mathbf{Z}^r_{l,o}-\mathbf{Z}^l_{r,o}}\right)\right]. \label{obj1app}
\end{alignat}
\end{subequations}

Recall that  $\{\mathbf{\Gamma}^r_l(t)\}$ and $\{\mathbf{\Lambda}^r_l(t)\}$, $\forall t, r, l$ are  symmetric  and skew-symmetric matrices, respectively. Therefore, $\text{Tr}((\mathbf{\Gamma}^r_l)^T \mathbf{Z}^r_l)$  and $\text{Tr}((\mathbf{\Lambda}^r_l)^T \mathbf{X}^r_l)$ are both equal to zeros.

We further define  $\mathbf{{N}}^r_{ij} = \frac{1}{2}(e^r_i{e^r_j}^T+e^r_j{e^r_i}^T)$ and $\mathbf{\hat{N}}^r_{ij}=  \frac{\sqrt{-1}}{2}(e^r_i{e^r_j}^T-e^r_j{e^r_i}^T)$, where $e_i^r$ is the $i$th basis vector in $\mathbb{R}^{\abs{\mathcal{B}_r}}$.
Therefore, Problem \eqref{SDP_OLD} is equivalent to 
\begin{subequations}\label{SDP_new}
\begin{alignat}{2}
&\argmin&\phantom{}&  \sum_{i=1}^{\abs{\mathcal{R}_r^g}}a_iP_{G_i}^2+b_iP_{G_i}+c_i + \sum_{l\in \mathcal{N}^{\delta{(r)}}}\left[\text{Tr}\left((\mathbf{\Gamma}^r_l + j\mathbf{\Lambda}^r_l)^H \right. \right. \notag\\
&\text{ }&\phantom{}& \left. \mathbf{W}^r_l \right)+  \alpha \left( \partial^{rl}_{k,i} \text{Tr} \{\mathbf{M}^r_i  \mathbf{W}^r\} + \partial^{rl}_{k,j} \text{Tr} \{\mathbf{M}^r_j  \mathbf{W}^r\} +\right. \notag\\
&\text{ }&\phantom{}& \left. \left. \partial^{rl}_{k, ij} \text{Tr} \{\mathbf{{N}}^r_{ij} \mathbf{W}^r \} +  \partial^{rl}_{k, {\hat{ij}}} \text{Tr} \{\mathbf{\hat{N}}^r_{{ij}} \mathbf{W}^r \} \right)\right] \label{obj2app}\\
&\text{s.t.}&\phantom{xx}& \text{ \eqref{SDPconst1}-\eqref{SDPconst_last1}. }
\end{alignat}
\end{subequations}
where $\partial^{rl}_{k,i}$, $\partial^{rl}_{k,j}$, $\partial^{rl}_{k, ij}$ and $\partial^{rl}_{k, {\hat{ij}}}$  denote  the subgradients   of $\abs{\mathbf{X}^r_{l,(i,i)}-\mathbf{X}^l_{r,(i,i)}}$, $\abs{\mathbf{X}^r_{l,(j,j)}-\mathbf{X}^l_{r,(j,j)}}$, $\abs{\mathbf{X}^r_{l,(i,j)}-\mathbf{X}^l_{r,(i,j)}}$ and $\abs{\mathbf{Z}^r_{l,(i,j)}-\mathbf{Z}^l_{r,(i,j)}}$ in the $k$th iteration, respectively. Note that buses $i,j$ are the boundary buses and these subgradients are the sign signals from the {system operator}. 

To transform the constraints to SDP, we also introduce some vectors:
\begin{enumerate}
	\item Group of active powers of generators:
\begin{equation}
\begin{split}\label{eq42}
	x_1=[P_{G_1}, d_{g_1}, \cdots, P_{G_i}, d_{g_i}, \cdots], \forall  i \in {{\mathcal{R}_r^g}}
\end{split}
\end{equation}
    \item Group of reactive powers of generators:
\begin{equation}
\begin{split}
	x_2=[Q_{G_1}, d_{g_1}, \cdots, Q_{G_i}, d_{g_i}, \cdots], \forall  i \in {{\mathcal{R}_r^g}}
\end{split}
\end{equation}
    \item Group of slack variables for  active powers:
   \begin{equation}
\begin{split}
	x_3=[u_{g_1}, l_{g_1}, \cdots, u_{g_i}, l_{g_i}, \cdots], \forall  i \in {{\mathcal{R}_r^g}}
\end{split}
\end{equation}
    \item Group of slack variables for  reactive powers:
 \begin{equation}
\begin{split}
	x_4=[u_{r_1}, l_{r_1}, \cdots, u_{r_i}, l_{r_i}, \cdots], \forall  i \in {{\mathcal{R}_r^g}}
\end{split}
\end{equation}
    \item Group of complex voltages:
 \begin{equation}
\begin{split}
	x_5=[V_{1},  \cdots, V_{i}, \cdots], \forall  i \in {{\mathcal{B}_r}}
\end{split}
\end{equation}
    \item Group of slack variables for complex voltages:
 \begin{equation}
\begin{split}
	x_6=[u_{b_1}, l_{b_1}, \cdots, u_{b_i}, l_{b_i}, \cdots], \forall  i \in {{\mathcal{B}_r}}\end{split}
\end{equation}
\end{enumerate}
Afterward, the SDP variable can be defined by:
 \begin{equation}
\begin{split}
	X=x^Hx,
\end{split}
 \end{equation}
where $x=[x_1, x_2, x_3, x_4, x_5, x_6]$. 
Therefore, the matrix $X$ is positive definite or semidefinite, i.e., $X \succeq 0$. {Note that the rank of $X$ should also be 1. A convex SDP relaxation is obtained by removing the rank constraint. By Theorem 9 of \cite{low2014convex}, when the cost function is convex and the network is a tree, the  SDP relaxation is exact under mild technical conditions. }
 \begin{equation}\label{bigx}
\begin{split}
X=
\begin{bmatrix}
X_{1}  & \cdots  & \cdots   & \cdots  & \cdots   & \cdots \\
\vdots & X_{2}   & \ddots   & \ddots  & \ddots   & \vdots \\
\vdots & \ddots  & X_{3}    & \ddots  & \ddots   & \vdots \\ 
\vdots & \ddots  & \ddots   & X_{4}   & \ddots   & \vdots \\
\vdots & \ddots  & \ddots   & \ddots  & X_{5}    & \vdots \\
\vdots & \cdots  & \cdots   & \cdots  & \cdots   & X_{6}\\
\end{bmatrix}.
\end{split}
 \end{equation}
Note that some elements in $X$ are replaced with ellipses, indicating that the relevant coefficients of those elements are always zeros. Therefore, the matrix $X$ can be treated as a block-diagonal symmetric matrix.  In particular, $X_5 = x_5^H  x_5  = (\mathbf{W}^r)^T$. We  define $X_1$, $X_2$, $X_3$, $X_4$ and  $X_6$   as follows.
 \begin{equation}
\begin{split}
X_{1}=
\begin{bmatrix}
P^2_{G_1}  & P_{G_1}d_{g_1}  & P_{G_1}P_{G_2}  & \cdots    & \cdots \\
P_{G_1}d_{g_1} & d^2_{g_1}   & \ddots   & \ddots     & \vdots \\
P_{G_1}P_{G_2}  & \ddots  & P^2_{G_2}    & \ddots     & \vdots \\ 
\vdots & \ddots  & \ddots   & d^2_{g_2}      & \vdots \\
\vdots & \cdots  & \cdots   & \cdots    & \cdots \\
\end{bmatrix}.
\end{split}
 \end{equation}
  \begin{equation}
\begin{split}
X_{2}=
\begin{bmatrix}
Q^2_{G_1}  & Q_{G_1}d_{g_1}  & Q_{G_1}P_{G_2}  & \cdots    & \cdots \\
Q_{G_1}d_{g_1} & d^2_{g_1}   & \ddots   & \ddots     & \vdots \\
Q_{G_1}P_{G_2}  & \ddots  & Q^2_{G_2}    & \ddots     & \vdots \\ 
\vdots & \ddots  & \ddots   & d^2_{g_2}      & \vdots \\
\vdots & \cdots  & \cdots   & \cdots    & \cdots \\
\end{bmatrix}.
\end{split}
 \end{equation}
 \begin{equation}
\begin{split}
X_{3}=
\begin{bmatrix}
u^2_{g_1}  & u_{g_1}l_{g_1}  & u_{g_1}u_{g_2}  & u_{g_1}l_{g_2}   & \cdots \\
u_{g_1}l_{g_1}  & l^2_{g_1}   & \ddots   & \ddots     & \vdots \\
u_{g_1}u_{g_2}  & \ddots  & u^2_{g_2}    & \ddots     & \vdots \\ 
u_{g_1}l_{g_2}  & \ddots  & \ddots   & l^2_{g_2}    & \vdots \\
\vdots & \cdots  & \cdots   & \cdots    & \cdots \\
\end{bmatrix}.
\end{split}
 \end{equation}
 \begin{equation}
\begin{split}
X_{4}=
\begin{bmatrix}
u^2_{r_1}  & u_{r_1}l_{r_1}  & u_{r_1}u_{r_2}  & u_{r_1}l_{r_2}   & \cdots \\
u_{r_1}l_{r_1}  & l^2_{r_1}   & \ddots   & \ddots     & \vdots \\
u_{r_1}u_{r_2}  & \ddots  & u^2_{r_2}    & \ddots     & \vdots \\ 
u_{r_1}l_{r_2}  & \ddots  & \ddots   & l^2_{r_2}    & \vdots \\
\vdots & \cdots  & \cdots   & \cdots    & \cdots \\
\end{bmatrix}.
\end{split}
 \end{equation}
  \begin{equation}
\begin{split}
X_{6}=
\begin{bmatrix}
u^2_{b_1}  & u_{b_1}l_{b_1}  & u_{b_1}u_{b_2}  & u_{b_1}l_{b_2}   & \cdots \\
u_{b_1}l_{b_1}  & l^2_{b_1}   & \ddots   & \ddots     & \vdots \\
u_{b_1}u_{b_2}  & \ddots  & u^2_{b_2}    & \ddots     & \vdots \\ 
u_{b_1}l_{b_2}  & \ddots  & \ddots   & l^2_{b_2}    & \vdots \\
\vdots & \cdots  & \cdots   & \cdots    & \cdots \\
\end{bmatrix}.
\end{split}
 \end{equation}

Ignoring the rank-1 constraint on $X$, problem \eqref{SDP_new} can be relaxed to the primal SDP  \eqref{SDP_prop1}, which is copied here for convenience:
\begin{subequations}\label{SDP_prop1_recall}
	\begin{alignat}{2}
    \text{Primal: }&\argmin \text{Tr}(A_0(\bm{\partial}))^T X), \\
    \text{s.t.  } & \text{Tr}(A_m^T X)=b_m, \forall m=\{1, \cdots, M\} \label{b_m} \\
     & X\succeq 0.
	\end{alignat}
	\end{subequations}
Here, the coefficient matrices $A_m$, $m=1, \cdots, M$ have the same dimension as $X$ and $\bm{\partial}$ is the set of all sign signals $\partial^{rl}_{k,i}$, $\partial^{rl}_{k,j}$, $\partial^{rl}_{k, ij}$ and $\partial^{rl}_{k, {\hat{ij}}}$, $\forall k, i, j, r, l$. The construction of the matrices $A_m$, $m=1, \cdots, M$,  are defined as follows:
 \begin{equation}
\begin{split}
A_m=
\begin{bmatrix}
A_{1, m}  & \cdots  & \cdots   & \cdots  & \cdots   & \cdots \\
\vdots    & A_{2, m}   & \ddots   & \ddots  & \ddots   & \vdots \\
\vdots    & \ddots  & A_{3, m}    & \ddots  & \ddots   & \vdots \\ 
\vdots & \ddots  & \ddots   & A_{4, m}   & \ddots   & \vdots \\
\vdots & \ddots  & \ddots   & \ddots  & A_{5, m}    & \vdots \\
\vdots & \cdots  & \cdots   & \cdots  & \cdots   & A_{6, m}\\
\end{bmatrix}.
\end{split}
 \end{equation}

In order to construct $A_0$, we first focus on the fuel cost of generator $G_i$ on  bus $i$, i.e., $\sum_{i=1}^{\abs{\mathcal{R}_r^g}}a_iP_{G_i}^2+b_iP_{G_i}+c_i$. Therefore,  $A_{1, 0}$ can be constructed as
 \begin{equation}
\begin{split}
A_{1, 0}=
\begin{bmatrix}
a_1  & 0.5b_1  & \cdots     & \cdots   &     \cdots  & \cdots   \\
0.5b_1     & c_1   & \ddots    & \ddots   & \vdots  & \cdots    \\
\vdots    & \ddots  & \ddots     & \ddots   & \vdots  & \cdots    \\ 
\vdots & \ddots  & \ddots    & a_i    & 0.5b_i  & \cdots   \\
\vdots & \cdots  & \cdots     & 0.5b_i   & c_i  & \cdots   \\
\vdots & \cdots  & \cdots    & \cdots   & \cdots  & \cdots   \\
\end{bmatrix}.
\end{split}
 \end{equation}

In addition, we have $\text{Tr}\left((\mathbf{\Gamma}^r_l + j\mathbf{\Lambda}^r_l)^H \mathbf{W}^r_l\right)$ for $ l\in \mathcal{N}^{\delta{(r)}}$, which can be constructed as 
 \begin{equation}
\begin{split}
A^{l, tr}_{5, 0}=  \mathbf{E}_{r\mapsto l}^T  (\mathbf{\Gamma}^r_l - j\mathbf{\Lambda}^r_l ) \mathbf{E}_{r\mapsto l}.
\end{split}
 \end{equation}

Similarly, we define $A^{l, (i,i)}_{5, 0} = \alpha \partial^{rl}_{k,i}  \{\mathbf{M}^r_i  \}$, $A^{l, (j,j)}_{5, 0} = \alpha \partial^{rl}_{k,j}  \{\mathbf{M}^r_j  \}$, $A^{l, (i,j)}_{5, 0} = \alpha \partial^{rl}_{k, ij} \{\mathbf{{N}}^r_{ij}  \}$, $ A^{l, (j,i)}_{5, 0} = \alpha \partial^{rl}_{k, {\hat{ij}}}  \{\mathbf{\hat{N}}^r_{{ij}}  \}$.

Therefore, $A_{5, 0}$ can be constructed as 
 \begin{equation}
\begin{split}
A_{5, 0} = \sum_{l\in \mathcal{N}^{\delta{(r)}}} \left(A^{l, tr}_{5, 0}+   \left[\sum_{o\in \mathcal{O}_{rl}} A^{l, o}_{5, 0} \right] \right).\label{A40}
\end{split}
 \end{equation}
In addition, $A_{2, 0}$, $A_{3, 0}$, $A_{4, 0}$, and $A_{6, 0}$ are zeros matrices.

In the following, we introduce the construction of  $A_i$ corresponding to the constraints. Here, we only consider \eqref{SDPconst1}, \eqref{SDPconst2}, \eqref{SDPconst3} and \eqref{SDPconst_last1}. Other constraints can be constructed in a similar way.

For  Eq. \eqref{SDPconst1} with $P_{G_1}$, the corresponding $A_m$ can be constructed as follows:
 \begin{equation}
\begin{split}
A_{5, m} = - {\mathbf{Y}}_i^r.
\end{split}
 \end{equation}
 In addition, we have
  \begin{equation}
\begin{split}
A_{1, m} = \begin{bmatrix}
\cdots      & 0.5     & \cdots     & \cdots    & \cdots  & \cdots   \\
0.5         & \ddots  & \ddots     & \ddots    & \vdots  & \cdots   \\
\vdots      & \ddots  & \ddots     & \ddots    & \vdots  & \cdots   \\ 
\vdots      & \ddots  & \ddots     & \ddots    & \cdots  & \cdots   \\
\vdots      & \cdots  & \cdots     & \cdots    & \cdots  & \cdots   \\
\vdots      & \cdots  & \cdots     & \cdots    & \cdots  & \cdots   \\
\end{bmatrix}.
\end{split}
 \end{equation}
Other terms of $A_{m}$, i.e., $A_{2, m}$, $A_{3, m}$, $A_{4, m}$, and $A_{6, m}$ corresponding to Eq. \eqref{b_m} are zeros matrices. Accordingly, $b_m$ in \eqref{b_m} is $P_{D_i}$.
 
 For   Eq. \eqref{SDPconst2} with $V_{1}$, the corresponding $A_m$ is constructed as follows.
 First of all, we have
  \begin{equation}
\begin{split}
A_{5, m} = {\mathbf{M}}_i^r,
\end{split}
 \end{equation}
 and
   \begin{equation}
\begin{split}
A_{6, m} = \begin{bmatrix}
1      & \cdots     & \cdots     & \cdots    & \cdots  & \cdots   \\
\cdots        & \ddots  & \ddots     & \ddots    & \vdots  & \cdots   \\
\vdots      & \ddots  & \ddots     & \ddots    & \vdots  & \cdots   \\ 
\vdots      & \ddots  & \ddots     & \ddots    & \cdots  & \cdots   \\
\vdots      & \cdots  & \cdots     & \cdots    & \cdots  & \cdots   \\
\vdots      & \cdots  & \cdots     & \cdots    & \cdots  & \cdots   \\
\end{bmatrix}.
\end{split}
 \end{equation}
In addition,  $A_{1, m}$, $A_{2, m}$, $A_{3, m}$, and $A_{4, m}$ corresponding to Eq. \eqref{SDPconst2} are zero matrices. Accordingly, $b_m$ in \eqref{b_m} is $\bar{V}^2_{1}$.
 
  For   Eq. \eqref{SDPconst3} with $P_{G_1}$, the corresponding $A_m$ is constructed as
   \begin{equation}
\begin{split}
A_{1, m} = \begin{bmatrix}
\cdots       & 0.5     & \cdots     & \cdots    & \cdots  & \cdots   \\
0.5        & \ddots  & \ddots     & \ddots    & \vdots  & \cdots   \\
\vdots      & \ddots  & \ddots     & \ddots    & \vdots  & \cdots   \\ 
\vdots      & \ddots  & \ddots     & \ddots    & \cdots  & \cdots   \\
\vdots      & \cdots  & \cdots     & \cdots    & \cdots  & \cdots   \\
\vdots      & \cdots  & \cdots     & \cdots    & \cdots  & \cdots   \\
\end{bmatrix},
\end{split}
 \end{equation}
 and
    \begin{equation}
\begin{split}
A_{3, m} = \begin{bmatrix}
1       & \cdots     & \cdots     & \cdots    & \cdots  & \cdots   \\
\cdots       & \ddots  & \ddots     & \ddots    & \vdots  & \cdots   \\
\vdots      & \ddots  & \ddots     & \ddots    & \vdots  & \cdots   \\ 
\vdots      & \ddots  & \ddots     & \ddots    & \cdots  & \cdots   \\
\vdots      & \cdots  & \cdots     & \cdots    & \cdots  & \cdots   \\
\vdots      & \cdots  & \cdots     & \cdots    & \cdots  & \cdots   \\
\end{bmatrix}.
\end{split}
 \end{equation}
Besides,   $A_{2, m}$, $A_{4, m}$, $A_{5, m}$, and $A_{6, m}$ corresponding to Eq. \eqref{SDPconst3} are zero matrices. Accordingly, $b_m$ in \eqref{b_m} is $\bar{P}_{G_1}$.
 
 Last, we give $A_m$ corresponding to  Eq. \eqref{SDPconst_last1} with $P_{G_1}$ as
     \begin{equation}
\begin{split}
A_{1, m} = \begin{bmatrix}
0      & \cdots     & \cdots     & \cdots    & \cdots  & \cdots   \\
\cdots       & 1  & \ddots     & \ddots    & \vdots  & \cdots   \\
\vdots      & \ddots  & \ddots     & \ddots    & \vdots  & \cdots   \\ 
\vdots      & \ddots  & \ddots     & \ddots    & \cdots  & \cdots   \\
\vdots      & \cdots  & \cdots     & \cdots    & \cdots  & \cdots   \\
\vdots      & \cdots  & \cdots     & \cdots    & \cdots  & \cdots   \\
\end{bmatrix}.
\end{split}
 \end{equation}
 Moreover,   $A_{2, m}$, $A_{3, m}$, $A_{4, m}$, $A_{5, m}$, and $A_{6, m}$ corresponding to Eq. \eqref{SDPconst_last1} are zero matrices and  $b_m$ in \eqref{b_m} is $1$. 
    
 This completes the proof of Proposition 1.
\hfill $\blacksquare$   

In the following, we show   the  equivalence between \eqref{obj1app} and \eqref{obj2app} if the subgradient KKT conditions are employed.

In order to construct $A'_{0}$, we formulate   $A'_{5, 0} $ as follows:
 \begin{equation}
\begin{split}
A'_{5, 0} = \sum_{l\in \mathcal{N}^{\delta{(r)}}} \left(A^{l, tr}_{5, 0} \right).
\end{split}
 \end{equation}
Other terms of  $A'_{0}$  are the same as those of $A_{0}$.

Because \eqref{obj1app} is non-differentiable, subdifferential versions of KKT conditions (Chapter 7 of \cite{ruszczynski2011nonlinear}) are:
\begin{subequations}
\begin{alignat}{2}
&\partial L_X= {A}'_0-\mu  X^{-1} - \sum_{m=1}^M A_m y_m - \sum_{l\in \mathcal{N}^{\delta{(r)}}} \bar{A}_0(\alpha \partial^r_{l}) ,\\
&\nabla L_y=b_m - \text{Tr}(A_m^T X)=0, \forall m=\{1, \cdots, M\}
\end{alignat}
\end{subequations}
where  $A_{5, 0}$ of $\bar{A}_0(\alpha \partial^r_{l})$  is denoted by 
\begin{equation}
\begin{split}
A_{5, 0}=&\alpha \partial^{rl}_{k,i}  \{\mathbf{M}^r_i\}+\alpha \partial^{rl}_{k,j}  \{\mathbf{M}^r_j \}+\alpha \partial^{rl}_{k, ij} \{\mathbf{{N}}^r_{ij}  \}+ \\
&\alpha \partial^{rl}_{k, {\hat{ij}}}  \{\mathbf{\hat{N}}^r_{{ij}}  \}.
\end{split}
\end{equation}
Recall that $A^{l, (i,i)}_{5, 0} = \alpha \partial^{rl}_{k,i}  \{\mathbf{M}^r_i  \}$, $A^{l, (j,j)}_{5, 0} = \alpha \partial^{rl}_{k,j}  \{\mathbf{M}^r_j  \}$, $A^{l, (i,j)}_{5, 0} = \alpha \partial^{rl}_{k, ij} \{\mathbf{{N}}^r_{ij}  \}$ and $ A^{l, (j,i)}_{5, 0} = \alpha \partial^{rl}_{k, {\hat{ij}}}  \{\mathbf{\hat{N}}^r_{{ij}}  \}$.  In addition, $\partial^{rl}_{k,i}= \pm 1$, $\partial^{rl}_{k,j} = \pm 1$, $\partial^{rl}_{k, ij}=\pm 1$ and $\partial^{rl}_{k, {\hat{ij}}}=\pm 1$. 
By incorporating all $\sum_{l\in \mathcal{N}^{\delta{(r)}}} \bar{A}_0(\alpha \partial^r_{l})$  into $A_0'$, we can obtain $A_0$. Therefore,  Problem \eqref{SDP_new} is equivalent to Problem \eqref{SDP_OLD}.



\subsection{Proof of Lemma 1}


For all non-zero $\mathbf{z}=\mathbf{x}+\text{j}\mathbf{y}$ in $\mathbb{C}^{N}$ with $\mathbf{x}$ and $\mathbf{y}$ in $\mathbb{R}^{N}$, we have
\begin{equation}\label{lem1}
\begin{split}
	&\mathbf{z}^H \mathbf{W} \mathbf{z} = (\mathbf{x}^T-\text{j} \mathbf{y}^T) (\mathbf{X} +\text{j}\mathbf{Z}) (\mathbf{x}+ \text{j} \mathbf{y})  \\
	&=\mathbf{x}^T \mathbf{X} \mathbf{x}+\mathbf{y}^T \mathbf{X} \mathbf{y} -  \mathbf{x}^T \mathbf{Z} \mathbf{y} + \mathbf{y}^T \mathbf{Z} \mathbf{x} \\
	&+\text{j} (\mathbf{x}^T \mathbf{Z} \mathbf{x}+\mathbf{y}^T \mathbf{Z} \mathbf{y} + \mathbf{x}^T \mathbf{X} \mathbf{y} -\mathbf{y}^T \mathbf{X} \mathbf{x} )
\end{split}
\end{equation}
It is obviously that $\mathbf{x}^T \mathbf{Z} \mathbf{x}=0$  for all non-zero $\mathbf{x}$ in $\mathbb{R}^{N}$ because $\mathbf{Z}$ is  skew-symmetric. Similarly, $\mathbf{y}^T \mathbf{Z} \mathbf{y}=0$. In addition, $\mathbf{x}^T \mathbf{X} \mathbf{y} -\mathbf{y}^T \mathbf{X} \mathbf{x} =0$ because $\mathbf{X}$ is  symmetric. Therefore, \eqref{lem1} is reduced into $\mathbf{x}^T \mathbf{X} \mathbf{x}+\mathbf{y}^T \mathbf{X} \mathbf{y} -  \mathbf{x}^T \mathbf{Z} \mathbf{y} + \mathbf{y}^T \mathbf{Z} \mathbf{x}$.
\begin{equation}\label{PSD}
\begin{split}
{\begin{bmatrix} \mathbf{x}  \\ \mathbf{y}  \end{bmatrix} }^T \begin{bmatrix} \mathbf{X} & -\mathbf{Z} \\ \mathbf{Z} & \mathbf{X} \end{bmatrix} {\begin{bmatrix} \mathbf{x}  \\ \mathbf{y}  \end{bmatrix} } = \mathbf{x}^T \mathbf{X} \mathbf{x}+\mathbf{y}^T \mathbf{X} \mathbf{y} - \mathbf{x}^T \mathbf{Z} \mathbf{y} + \mathbf{y}^T \mathbf{Z} \mathbf{x}.
\end{split}
\end{equation}
Therefore, the semi-definiteness of \eqref{lem1} and \eqref{PSD} can imply each other.

This completes the proof.
\hfill $\blacksquare$
 
\begin{footnotesize}

\end{footnotesize}

\begin{IEEEbiography}
[{\includegraphics[width=1in,height=1.25in,clip,keepaspectratio]{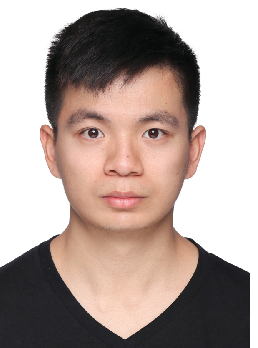}}]
{Tong Wu} (Student Member, IEEE)  is currently working toward the Ph.D. degree in the Department of Information Engineering, The Chinese University of Hong Kong, Hong Kong. His research interests include machine learning and distributed optimization in smart grids. He is also interested in  graph representation learning and statistical learning.
\end{IEEEbiography}

\begin{IEEEbiography}
[{\includegraphics[width=1in,height=1.25in,clip,keepaspectratio]{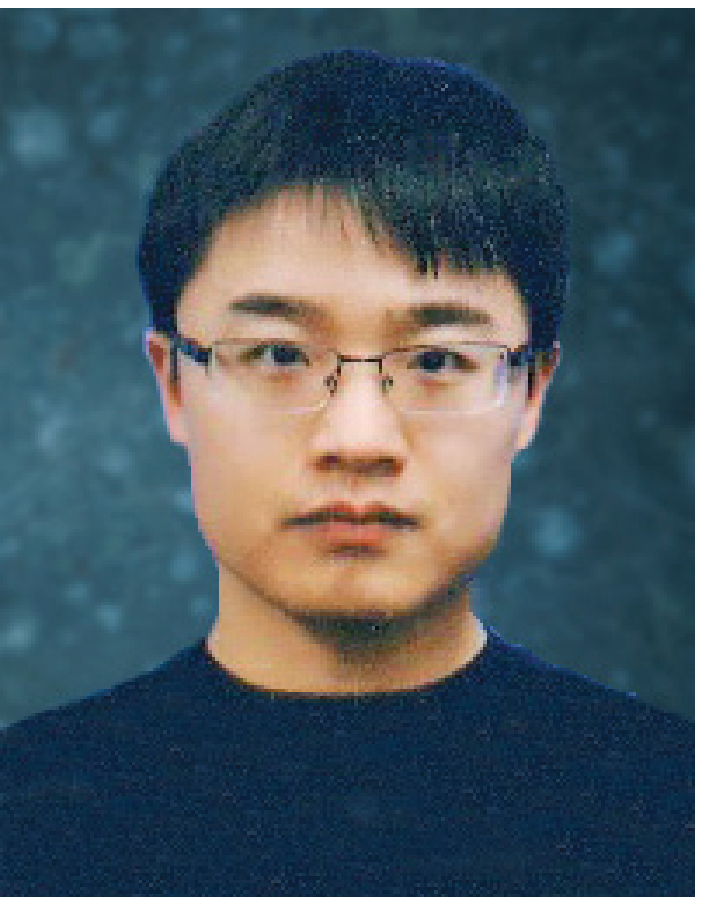}}]
{Changhong Zhao} (Member, IEEE) is an Assistant Professor with the Department of Information Engineering, the Chinese University of Hong Kong. He received B.E. degree in automation from Tsinghua University, Beijing, China, in 2010, and M.S. and Ph.D. degrees in electrical engineering from California Institute of Technology, Pasadena, CA, USA, in 2012 and 2016, respectively. From 2016 to 2019, he worked with the National Renewable Energy Laboratory, Golden, CO, USA. His research is on control and optimization of network systems such as smart grid. He received Demetriades Prize and Wilts Prize for best Ph.D. thesis from Caltech EAS Division and EE Department, respectively, and Early Career Award from Hong Kong Research Grants Council.
\end{IEEEbiography}

\begin{IEEEbiography}
[{\includegraphics[width=1in,height=1.25in,clip,keepaspectratio]{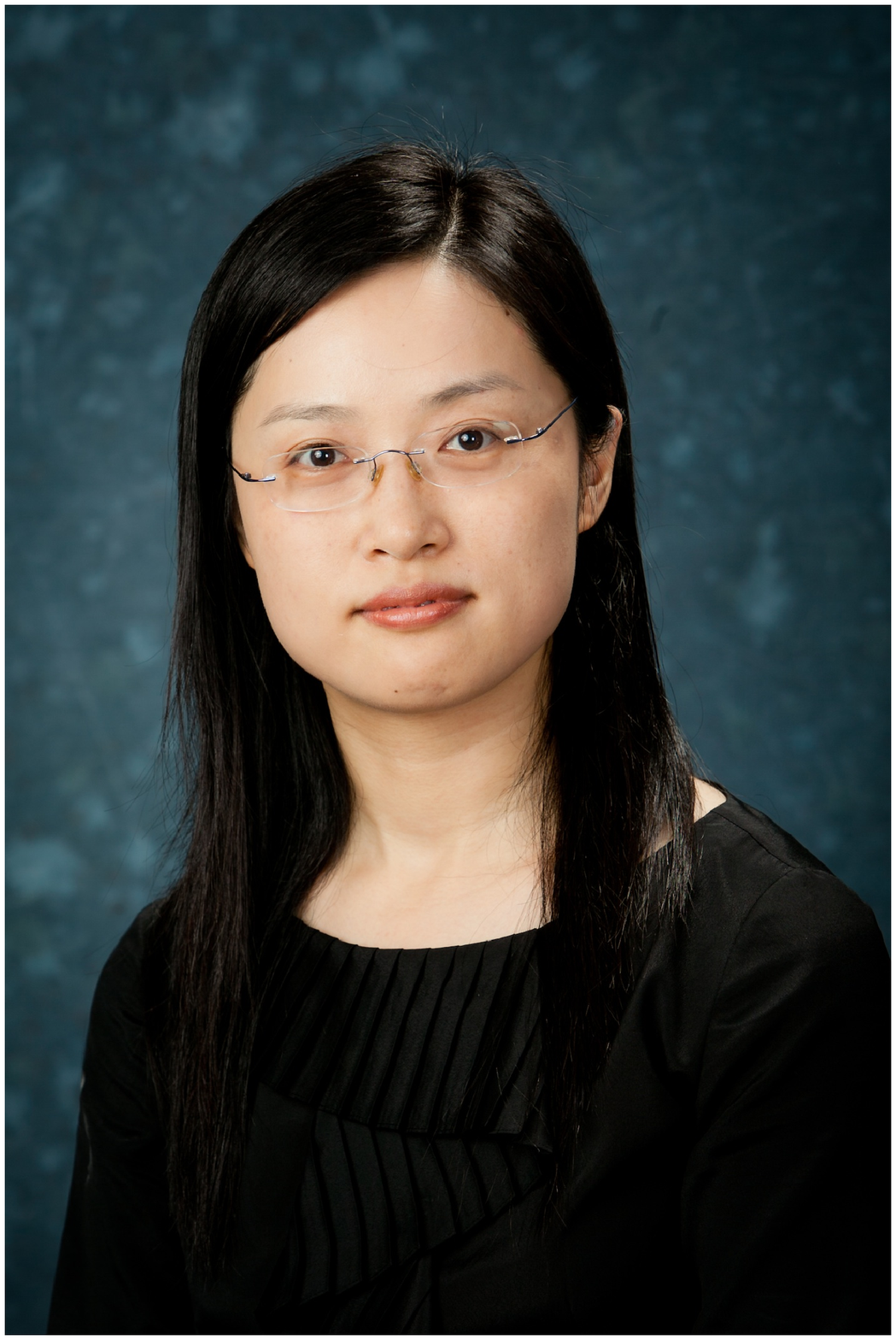}}]
{Ying-Jun Angela Zhang} (Fellow, IEEE)  is currently a Professor at The Chinese University of Hong Kong. She received her Ph.D. degree from The Hong Kong University of Science and Technology. She has published actively in the area of wireless communication systems and smart power grid, and received several Best Paper Awards, including the IEEE Marconi Prize Paper Award in Wireless Communications. She is a Fellow of IET, IEEE and a Distinguished Lecturer of IEEE ComSoc.

She is currently a Member of IEEE ComSoc Fellow Evaluation Committee and an Associate Editor-in-Chief of IEEE Open Journal of the Communications Society. She is on the Steering Committees of IEEE Wireless Communications Letters and IEEE SmartgridComm Conference. Previously, she served as the Chair of Executive Editorial Committee of IEEE Transactions on Wireless Communications, and the Founding Chair of IEEE Smart Grid Communications Technical Committee. She has been a Technical Program/Symposium Co-Chair of IEEE ICC/GC  Conferences multiple times.
\end{IEEEbiography}

\end{document}